\documentclass[12pt]{article}
\usepackage{amsfonts,amsmath}
\usepackage{amssymb}

\usepackage[hypertex] {hyperref}

\newcommand{\Hp}{\Sp^+}

\newcommand{\by}{\bar{y}}
 
\newcommand{\T}{\mathcal{T}}
\newcommand{\Sp}{{\mathcal H}}

\newcommand{\dr}{{{\rm d}}}

\renewcommand{\theequation}{\thesection.\arabic{equation}}
\renewcommand{\thesubsection}{\arabic{section}.\arabic{subsection}}
\makeatletter \@addtoreset{equation}{section} \makeatother

\newcommand{\gt}{\tau}
 \newcommand{\eq}{\eqref}
\newcommand{\ie}{{\it i.e.,} }

\newcommand{\rhs}{{\it r.h.s.} }
\newcommand{\rhss}{{\it r.h.s.'s} }

\newcommand{\be}{\begin{equation}}
\newcommand{\ee}{\end{equation}}
\newcommand{\bee}{\begin{eqnarray}}
\newcommand{\beee}{\begin{array}}
\newcommand{\eee}{\end{eqnarray}}
\newcommand{\eeee}{\end{array}}
\newcommand{\gn}{\nu}

\newcommand{\gx}{\xi}
\newcommand{\gr}{\rho}
\newcommand{\ga}{\alpha}
\newcommand{\gb}{\beta}

\newcommand{\gga}{\gamma}

\newcommand{\ls}{\!\!\!\!\!\!}

\newcommand{\gd}{\delta}

\newcommand{\gvep}{\varepsilon}
\newcommand{\gs}{\sigma}
\newcommand{\go}{\omega}

\newcommand{\q}{\,,\qquad}
\newcommand{\nn}{\nonumber}
\newcommand{\half}{\frac{1}{2}}

\newcommand{\p}{\partial}

\newcommand{\ff}{\frac}

\newcommand{\II}{{\cal I}}
\newcommand{\Ee}{{\cal E}}
\newcommand{\Ez}{{E_z{\,}}}
\newcommand{\Pz}{\mathbb{P}}
\newcommand{\PP}{{\cal P}}
\newcommand{\chalf}{\frac{1}{4}}
\newcommand{\KE}{{\mathsf{F}}}
\newcommand{\KEE}{{\mathbb{F}}}

\begin{document}

\begin{flushright}
FIAN/TD/21-2020\\
\end{flushright}

\vspace{0.5cm}
\begin{center}
{\large\bf Manifest Form   of  the Spin-Local
Higher-Spin Vertex $\Upsilon^{\eta\eta}_{\go CCC}$}

\vspace{1 cm}

\textbf{ O.A.~Gelfond${}^{1,2}$ and A.V.~Korybut${}^1$ }\\

\vspace{1 cm}

\textbf{}\textbf{}\\
 \vspace{0.5cm}
 \textit{${}^1$ I.E. Tamm Department of Theoretical Physics,
Lebedev Physical Institute,}\\
 \textit{ Leninsky prospect 53, 119991, Moscow, Russia }\\

\vspace{0.7 cm}\textit{
${}^2$ Federal State Institution "Scientific Research Institute for System Analysis
of the Russian Academy of Science",}\\
\textit{Nakhimovsky prospect 36-1, 117218, Moscow, Russia\, }

\par\end{center}

\begin{center}

\vspace{0.7 cm}   gel@lpi.ru, akoribut@gmail.com
\par\end{center}

\vspace{0.4cm}

\begin{abstract}
\noindent
{Vasiliev generating system of higher-spin equations allowing to reconstruct nonlinear vertices of field equations for higher-spin gauge fields contains a free complex parameter $\eta$. Solving the generating system order by order one obtains physical vertices proportional to various powers of $\eta$ and $\bar{\eta}$. Recently  $\eta^2$ and $\bar{\eta}^2$ vertices  in the zero-form sector were presented
  in \cite{4a3} in the $Z$-dominated form implying  their spin-locality by virtue of $Z$-dominance Lemma of \cite{2a1}. However the vertex of  \cite{4a3} had
  the form of a sum of spin-local terms dependent on the auxiliary spinor variable $Z$ in the
  theory modulo so-called $Z$-dominated terms, providing a sort of existence theorem
  rather than explicit form of the vertex. The aim of this paper is to elaborate
  an approach allowing to systematically  account for the effect of $Z$-dominated terms
  on the final $Z$-independent form of the vertex needed for any practical analysis.
  Namely, in this paper we obtain explicit $Z$-independent spin-local  form  for
  the vertex  $\Upsilon^{\eta\eta}_{\go CCC}$ for its $\go CCC$-ordered part
  where $\go$ and $C$ denote gauge one-form and field strength zero-form higher-spin fields valued in an
   arbitrary associative algebra in which case the order of product  factors in the vertex matters.   The developed formalism is based on the Generalized Triangle identity derived in the paper and  is applicable to all other
orderings of the fields in the vertex.   }\end{abstract}

\newpage

\tableofcontents

\section{Introduction}

Higher-spin (HS) gauge theory describes interacting systems of massless fields of all spins
(for reviews see e.g. \cite{{Vasiliev:Golfandmem},Review4}).
 Effects of HS gauge theories are anticipated to play a role at ultra high energies of
Planck scale \cite{Vasiliev:2016xui}.
Theories of this class play a role in various contexts from holography \cite{Klebanov:2002ja} to cosmology
\cite{Barv}. HS theory
differs from usual local field theories because it contains
infinite tower of gauge fields of all spins and the number of space-time derivatives increases with the spins
of fields  in the vertex \cite{Bengtsson:1983pd,Berends:1984wp,Fradkin:1987ks,Fradkin:1991iy}.
However one may ask for spin-locality \cite{Vasiliev:2016xui,Gelfond:2017wrh,4a1,4a2} which implies
space-time locality in the lowest orders of perturbation theory \cite{4a1}. Even though details of the precise relation between spin-locality and space-time locality in higher orders of perturbation theory have not been yet elaborated,  from the form of equations it is clear that spin-locality constraint provides one of the best tools to minimize the space-time non-locality.  Moreover demanding spin-locality one actually fixes functional space for possible field redefinitions that is highly important for
the predictability of the theory.

A useful way of description of HS dynamics is provided by the
generating Vasiliev system of HS equations  \cite{more}. The latter {contains a free complex parameter
 $\eta$. Solving the generating system order by order one obtains vertices
proportional to various powers of $\eta$ and $\bar{\eta}$. In the recent paper \cite{4a3},
$\eta^2$ and $\bar{\eta}^2$  vertices were obtained in the sector of equations for zero-form fields,
containing, in particular, a part of the $\phi^4$ vertex for the scalar field $\phi$
  in the theory. Though being seemingly $Z$-dependent, in \cite{4a3} these vertices were written
  in the $Z$-dominated form which implies  their spin-locality by virtue of $Z$-dominance
  Lemma of \cite{2a1}. In this paper we obtain explicit $Z$-independent spin-local  form  for
  the vertex  $\Upsilon^{\eta\eta}_{\go CCC}$ starting from the $Z$-dominated expression of \cite{4a3}.
  The label $\go CCC$ refers to the $\go CCC$-ordered part of the vertex
  where $\go$ and $C$ denote gauge one-form and field strength zero-form HS fields valued
  in arbitrary associative algebra in which case the order of the product  factors in $\go CCC$ matters.
}

There are several  ways to study the issue of (non)locality in HS gauge theory. One is reconstruction
the vertices from the boundary by the  holographic prescription based on the Klebanov-Polyakov
 conjecture \cite{Klebanov:2002ja} (see also
\cite{Sezgin:2002rt}, \cite{SS}). Alternatively,
one can  analyze vertices  directly in the bulk starting from the generating equations
of \cite{more}. The latter approach developed in \cite{4a1,4a2,4a3,2a1,2a2}
is free from  any holographic duality  assumptions but demands careful choice of the
homotopy scheme to determine the choice of field variables compatible with spin-locality of the vertices.
The issue of (non)locality of HS gauge theories was also considered in
\cite{Fotopoulos:2010ay} and \cite{David:2020ptn} with somewhat opposite conclusions.

From the holographic point of view the vertex that contains $\phi^4$ was argued to be essentially
non-local \cite{Sleight:2017pcz} or at least should have  non-locality of very specific form presented
in \cite{Ponomarev:2017qab}. On the other hand, the holomorphic, \ie
$\eta^2$ and antiholomorphic $\bar\eta^2$ vertices, where $\eta$ is a complex parameter in
the HS equations, were recently obtained in \cite{4a3} where they were
shown to be spin-local by virtue of $Z$-dominance lemma of \cite{2a1}.
The computation was done directly in the bulk starting from the non-linear HS system of \cite{more}.

In this formalism HS fields  are described by one-forms
$\omega (Y;K|x) $ and zero-forms $C(Y;K|x)$ where $x$ are space-time coordinates while
$Y_A=(y_\ga,\by_{\dot \ga})$ are auxiliary spinor variables.
Both dotted and undotted indices are two-component, $\ga, {\dot{\ga}=1,2}$, while $K=(k,\bar k)$ are outer
Klein operators satisfying $k*k= \bar{k}* \bar{k}=1$\,,
 \bee\label{Klein}&&
\lbrace k,y^\ga\rbrace_\ast=\lbrace k, z^\ga \rbrace_\ast=
\lbrace \bar{k},\bar{y}^{\dot{\ga}}\rbrace_\ast=\lbrace \bar{k},\bar{z}^{\dot{\ga}}
\rbrace_\ast=\lbrace k,\theta^\ga\rbrace_\ast=\lbrace \bar{k},\bar{\theta}^{\dot{\ga}}\rbrace_\ast=0,
\\\nn&&[k,\bar{y}^{\dot{\ga}}]_\ast=[ k, \bar{z}^{\dot{\ga}}]_\ast=
[\bar{k},y^\ga]_\ast=[ \bar{k},z^\ga]_\ast=[k,\bar{\theta}^{\dot{\ga}}]_\ast=[ \bar{k},\theta^\ga]_\ast=0\,,
\eee
where $\theta $ and $\bar \theta $ are anticommuting spinors in the theory.

Schematically, non-linear HS equations in the unfolded form read as
\begin{equation}\label{oneform}
\dr_x \go + \go \ast \go=\Upsilon(\go,\go,C)+\Upsilon(\go,\go,C,C)+\ldots,
\end{equation}
\begin{equation}\label{zeroform}
\dr_x C+\go \ast C-C\ast \go=\Upsilon(\go,C,C)+\Upsilon(\go,C,C,C)+\ldots.
\end{equation}

As recalled in Section \ref{HSeq}, generating equations of \cite{more} that reproduce the form of
equations (\ref{oneform}) and (\ref{zeroform}) have a simple form as a result of
doubling of spinor variables,
namely $$
\go(Y;K|x)\longrightarrow W(Z;Y;K|x)\,,\qquad
C(Y;K|x) \longrightarrow B(Z;Y;K|x). $$
Equations
\eq{oneform} and \eq{zeroform} result from the generating equations of \cite{more} upon
order by order reconstruction of $Z$-dependence  (for more detail see Section \ref{HSeq}).
The final form of equations (\ref{oneform}) and (\ref{zeroform}) turns out
to be  $Z$-independent  as a consequence of consistency of the  equations of \cite{more}. This fact may not be
manifest however since the \rhss of HS equations
usually have the form of the sum of $Z$-dependent terms.

HS equations  have remarkable property  \cite{Vasiliev:1988sa} that they remain
consistent with the fields $W$ and $B$ valued in any associative algebra. For instance
$W$ and $B$ can belong to the matrix algebra $Mat_n $ with any $n$. Since in that
case the components of $W$ and $B$ do not commute, different orderings of the fields
should be considered independently.
(Mathematically, HS equations with this property correspond to $A_\infty $ strong homotopy
 algebra introduced by Stasheff in \cite{stash1},\cite{stash2},\cite{stash3}.)
For instance, holomorphic (\ie $\bar\eta$-independent) vertices in the zero-form sector can be represented in the form
\be \label{FieldOrdering}
\Upsilon^{\eta }(\go,C,C )=\Upsilon^{ \eta}_{\go CC }+\Upsilon^{ \eta}_{C\go  C }+\Upsilon^{ \eta}_{CC\go}
\,,\quad
\Upsilon^{\eta\eta}(\go,C,C,C)=\Upsilon^{\eta\eta}_{\go CCC}+\Upsilon^{\eta\eta}_{C\go CC}
+\Upsilon^{\eta\eta}_{CC\go C}+\Upsilon^{\eta\eta}_{CCC\go }\,,\,\,\ldots
\ee
where the subscripts of the vertices $\Upsilon$ refer to the ordering of the product
factors.

The vertices obtained in \cite{4a3}  were shown to be
spin-local due to  the $Z$-dominance Lemma of \cite{2a1} that identifies terms that
must drop from the \rhss of HS equations together with the $Z$-dependence.
Recall that spin-locality implies that the vertices are local in terms of spinor variables for
any finite subset of fields of different spins \cite{2a2} (for more detail on the notion of spin-locality see \cite{2a2}).
    Analogous  vertices in the one-form sector have been  shown to be spin-local earlier in \cite{4a2}.

 The main achievement of \cite{4a3} consists of finding such  solution of the generating
 system in the third order in  $C$  that all  spin-nonlocal terms  containing infinite towers
 of derivatives in $y(\bar y)$  between $C$-fields in the (anti)holomorphic
 in $\eta(\bar \eta)$ sector do not contribute to
 $\eta^2$ ($\bar \eta^2$) vertices by virtue of
 $Z$-dominance Lemma. Thus \cite{4a3} gives  spin-local expressions for the vertices
 $\Upsilon^{\eta\eta}(\go,C,C,C)$ which, however, have a form
 of a sum of a number of  $Z$-dependent terms. To make spin-locality
 manifest one must remove the seeming Z-dependence from the vertex of \cite{4a3}.
 Technically, this can be done with the help of partial integration  and  the Schouten identity.
  The aim of this paper is to show how this works in practice.

Since the straightforward  derivation presented in this paper
is technically involved we confine ourselves to the
particular vertex $\Upsilon^{\eta\eta}_{\go CCC}$ \eq{FieldOrdering}.
{      Complexity of the calculations in this paper expresses complexity of the
obtained vertex  having no analogues in the literature. Indeed, this is explicitly calculated
spin-local vertex
of the third order in the equations, corresponding to the vertices
of the fourth (and, in part, fifth) order for the fields of all spins.
The example described in the paper explains the formalism applicable to all other orderings
of the fields in the vertex that are also computable. So, our results are most important from the general
 point of view
  highlighting  a way for the computation of higher vertices in HS theory that may
  be important from various perspectives and, in the first place, for the analysis of HS holography.
It should be stressed that the results of \cite{4a3} provided a sort of existence theorem for a spin-local
vertex  that was difficult to extract without developing specific tools like those developed in
this paper.
In particular, it is illustrated how the general statements
like $Z$-dominance Lemma work in practical computations. Let us stress that
at the moment this is the only available approach allowing to compute explicit form
of the spin-local vertices for all spins   at higher orders.}

The rest of the paper is organized as follows. In Section \ref{HSeq}, the necessary background on HS equations
is presented with brief recollection on the procedure of derivation of vertices from the generating system.
 Section \ref{SectionHplus}
reviews  the notion of the $\Hp$ space as well as the justification for a computation modulo $\Hp$.
In Section \ref{Schema}, we present step-by-step scheme of computations performed in this paper.
 Section \ref{Main} contains the final
manifestly spin-local expression for  $\Upsilon^{\eta\eta}_{\go CCC}$ vertex.
In Sections \ref{zlinear}\,, \ref{SecGTid}\,, \ref{uniform}\,, \ref{Eli0} and \ref{proof}
technical details of the steps
sketched in Section  \ref{Schema} are presented. In particular, in Section \ref{SecGTid}
we introduce   important {\it Generalised Triangle identity} which allows us to uniformize expressions
from \cite{4a3}.
 Conclusion section contains discussion of the obtained results.  Appendices A, B, C  and D contain
 technical detail on the   steps listed in
the scheme of computation.
 Some useful formulas are collected in Appendix E.

\section{ Higher Spin equations}
\label{HSeq}
\subsection{Generating   equations}

Spin-$s$ HS fields are encoded in two generating functions, namely, the space-time one-form
\begin{equation}
\omega(y,\bar{y},x)=\dr x^\mu \go_\mu(y,\bar{y},x)=\sum_{n,m} \dr x^{\mu} \go_{\mu} {}_{\ga_1 \ldots \ga_n,
 \dot{\ga}_1 \ldots \dot{\ga}_m}(x) y^{\ga_1} \ldots y^{\ga_n} \bar{y}^{\dot{\ga}_1}
 \ldots \bar{y}^{\dot{\ga}_m}
\q s=\ff{2+m+n}{2} \end{equation} and  zero-form
\begin{equation}
C(y,\bar{y},x)=\sum_{n,m} C_{\ga_1 \ldots \ga_n,
\dot{\ga}_1 \ldots \dot{\ga}_m}(x) y^{\ga_1} \ldots y^{\ga_n}
\bar{y}^{\dot{\ga}_1} \ldots \bar{y}^{\dot{\ga}_m}
\q   s=\ff{|m-n|}{2}.\end{equation}
where $\ga=1,2$ and  $\dot{\ga}=1,2$ are two-component spinor indices.
 Auxiliary commuting variables $y^\ga$ and $\bar{y}^{\dot \ga}$
 can be combined into an $\mathfrak{sp}(4)$ spinor $Y^A=(y^\ga,\bar{y}^{\dot{\ga}})$, $A=1,..., 4$.

The vertices  $\Upsilon(\go,\go,C,C,\ldots)$  \eqref{oneform} and  $\Upsilon(\go,C,C,\ldots)$
\eqref{zeroform} result from
 the generating system of \cite{more}
\begin{equation}\label{HS1}
\dr_x W+W\ast W=0,
\end{equation}
\begin{equation}\label{HS2}
\dr_x S+W\ast S+S\ast W=0,
\end{equation}
\begin{equation}\label{HS3}
\dr_x B+W\ast B- B\ast W=0,
\end{equation}
\begin{equation}\label{HS4}
S\ast S=i(\theta^A \theta_A+\eta B\ast \gga+\bar{\eta} B\ast \bar{\gga}),
\end{equation}
\begin{equation}\label{HS5}
S\ast B-B\ast S=0.
\end{equation}
Apart from space-time coordinates $x$, the fields $W(Z;Y;K|x)$, $S(Z;Y;K|x)$ and $B(Z;Y;K|x)$
 depend on $Y^A$, $Z^A=(z^\ga,\bar{z}^{\dot{\ga}})$ and Klein operators $K=(k,\bar{k})$
\eq{Klein}. $W$ is a space-time  one-form, \ie $W= dx^\nu W_\nu$
while  $S$ -field is a one-form in $Z$ spinor directions
$\theta^A=(\theta^\ga,\bar{\theta}^{\dot{\ga}})$,\quad $\lbrace\theta^A, \theta^B\rbrace=0$, \ie
\begin{equation}
S(Z;Y;K)=\theta^A S_A(Z;Y;K).
\end{equation}
$B$ is a zero-form.

Star product is defined as follows
\begin{equation}\label{StarZY}
(f\ast g)(Z;Y;K)=\frac{1}{(2\pi)^4}\int d^4 U \, d^4 V e^{iU_A V^A}f(Z+U,Y+U;K)g(Z-V,Y+V;K).
\end{equation}
  Elements
 \begin{equation}
\gga=\theta^\ga \theta_\ga e^{iz_\ga y^\ga}k\mbox{\qquad and\qquad}
\bar{\gga}=\bar{\theta}^{\dot{\ga}}\bar{\theta}_{\dot{\ga}}
e^{i\bar{z}_{\dot{\ga}}\bar{y}^{\dot{\ga}}}\bar{k}
\end{equation}
are central because $\theta^3=0$ since $\theta_\ga$ is a two-component anticommuting spinor.
\subsection{Perturbation theory}
Starting with a particular solution of the form
\begin{equation}\label{solution}
B_0(Z;Y;K)=0\q S_0(Z;Y;K)=\theta^\ga z_\ga+\bar{\theta}^{\dot{\ga}}\bar{z}_{\dot{\ga}}\q
 W_0(Z;Y;K)=\omega(Y;K)\,,
\end{equation}
which  indeed  solves  \eqref{HS1}-\eqref{HS5} provided that  $\go(Y;K)$ satisfies zero-curvature condition,
\be
\dr \go +\go*\go=0\,,
\ee
 one develops perturbation theory. Starting from \eqref{HS5} one finds
\begin{equation}\label{1order}
[S_0,B_1]_*=0.
\end{equation}
From \eqref{StarZY} one  deduces that
\begin{equation}
[Z_A,f(Z;Y;K)]_\ast=-2i\frac{\p}{\p Z^A} f(Z;Y;K).
\end{equation}
Hence, equation (\ref{1order}) yields
\begin{equation}
[S_0,B_1]=-2i \theta^A \frac{\p}{\p Z^A}B_1=-2i \dr_Z B_1=0 \; \Longrightarrow\; B_1(Z;Y;K)=C(Y;K).
\end{equation}
The $Z$-independent $C$-field that appears as the first-order part of $B$ is the same
 that enters equations \eqref{oneform}, \eqref{zeroform}. The perturbative procedure can be
 continued further leading to the equations of the form
\begin{equation}
\dr_Z \Phi_{k+1}=J(\Phi_k, \Phi_{k-1},\ldots)\,,
\end{equation}
where $\Phi_k$ is either $W$, $S$ or $B$ field of the $k$-th order of perturbation theory,
 identified with the degree of $C$-field in the corresponding expression, \ie
\bee\nn&&
W=\go+W_1(\go,C)+W_2(\go,C,C)+\ldots\q S=S_0+S_1(C)+S_2(C,C)+\ldots,
\\&&\nn B=C+B_2(C,C)+B_3(C,C,C)+\ldots.
\eee
To obtain dynamical equations \eqref{oneform}, \eqref{zeroform} one should plug obtained
 solutions into equations \eqref{HS1} and \eqref{HS3}. For instance,
 \eqref{HS3} up to the third order in $C$-field is
\begin{equation}\label{B3EQ}
\dr_x C+[\go,C]_\ast=-\dr_x B_2-[W_1,C]_\ast-\dr_x B_3-[W_1,B_2]_\ast-[W_2,C]_\ast+\ldots
\end{equation}
Though the fields $W_1$, $W_2$ and $B_2$, $B_3$ and hence various terms that enter   (\ref{B3EQ})
are  $Z$-dependent, equations \eqref{HS1}-\eqref{HS5} are designed in such a way that, as a consequence
of their consistency, the sum of the terms on the \rhs
of (\ref{B3EQ}) is $Z$-independent. To see this it suffices to apply $\dr_Z$
realized as $\ff{i}{2} [S_0\,,\quad ]_*$ to the \rhs
of (\ref{B3EQ}) and make sure  that it gives zero by virtue of already solved equations.
For more detail we refer the reader to the review \cite{Review4}.

\section{ Subspace $\Hp$ and $Z$-dominance lemma}
\label{SectionHplus}

\subsection{$\Hp$}

In this Section the definition of the      space $\Hp$ \cite{4a3} that plays a
crucial role in our computation is recollected.
Function $f(z,y\vert \theta)$  of the form
\begin{equation}
\label{class}
f(z,y\vert \theta)=\int_0^1 d\mathcal{T}\, e^{i\mathcal{T}z_\ga y^\ga}\phi
\left(\mathcal{T}z,y\vert \mathcal{T} \theta,\mathcal{T}\right)\,
\end{equation}
 belongs to the space $\Hp$ if there exists
 such a real $\varepsilon>0$, that
\begin{equation}\label{limit}
\lim_{\mathcal{T}\rightarrow 0}\mathcal{T}^{1-\varepsilon}\phi(w,u\vert
\theta,\mathcal{T})=0\,.
\end{equation}
Note that this definition  does not demand
any specific behaviour of  $\phi$ at $\mathcal{T}\to1$ as was the case for the
 space $\Sp^{+0}$ of \cite{2a2}.

 In the sequel we   use  two main types of functions that obey \eqref{limit}:
\begin{equation}\label{kernels}
\phi_1(\mathcal{T}z,y\vert \mathcal{T} \theta,
\mathcal{T})=\frac{\mathcal{T}^{\delta_1}}{\mathcal{T}}\widetilde{\phi}_1(\mathcal{T}z,y\vert
\mathcal{T} \theta)\q \phi_2(\mathcal{T} z,y\vert
\mathcal{T}\theta,
\mathcal{T})=\vartheta(\mathcal{T}-\delta_2)\frac{1}{\mathcal{T}}\widetilde{\phi}_2(\mathcal{T}z,y\vert
\mathcal{T} \theta)
\end{equation}
with some $\delta_{1,2}>0$. (Note that the second option with $\delta_2>0$ can be
interpreted as the first one with arbitrary large $\delta_1$. Here   step-function is denoted as $\vartheta$
to distinguish it from the anticommuting variables $\theta$.)

Space $\Hp$ can be represented as the direct sum
\begin{equation}
\Hp=\Hp_0 \oplus \Hp_1 \oplus \Hp_2\,,
\end{equation}
where $\phi(w,u\vert\theta,\mathcal{T})\in\Hp_p$ are degree-$p$ forms in  $\theta$ satisfying \eqref{limit}.

All terms from  $\Hp$ on the \rhs of HS field equations  must vanish by $Z$-dominance Lemma \cite{2a1}.
Following \cite{4a3} this can be understood as follows. All the expressions
from \eqref{B3EQ} have the form  \eqref{class} and the only way to obtain $Z$-independent non-vanishing
 expression is to bring the hidden $\T$ dependence in $\phi(\T z,y\vert \T \theta , {\T})$
 to $\delta(\T)$. If a function contains an additional factor of
$\mathcal{T}^\gvep$ or is isolated  from $\T=0$, it cannot contribute to the  $Z$-independent
answer
 which is the content of $Z$-dominance Lemma \cite{2a1}.
This just means that functions of the class $\Hp_0$ cannot
contribute to the $Z$-independent equations  \eqref{zeroform}.
Application of this fact to locality is straightforward once this is
shown that all terms containing
infinite towers of higher derivatives in the vertices of interest
belong to $\Hp_0$ and, therefore, do not contribute to HS
equations. This is what was in particular shown  in \cite{4a3}.

\subsection{Notation}
As in \cite{4a3} we   use \textit{exponential} form for all the expressions below where by $\go CCC$ we assume
\begin{equation}
\omega(\mathsf{y}_\go,\bar{y})\bar{\ast}C(\mathsf{y}_1,\bar{y})\bar{\ast}C(\mathsf{y}_2,\bar{y})\bar{\ast}C(\mathsf{y}_3,\bar{y})
\end{equation}
with $\bar{\ast}$ denoting star-product with respect to $\bar{y}$.
Derivatives $\p_\go$ and $\p_j$ act on auxiliary variables as follows
\begin{equation}
\p_{\go\ga}=\frac{\p}{\p \mathsf{y}_\go^\ga}\q \p_{j\ga}=\frac{\p}{\p \mathsf{y}_j^\ga}.
\end{equation}
After all the derivatives   in $\mathsf{y}_\go$ and $\mathsf{y}_j$ are evaluated the latter are set to zero, \ie
\begin{equation}
\mathsf{y}_\go=\mathsf{y}_j=0.
\end{equation}
In this paper we use the following  notation of \cite{4a3}:
\begin{equation}
t_\ga:=-i\p_{\go\ga},\;\; p_{j\ga}:=-i\p_{j\ga}\q
\end{equation}
 \be\label{ro+}
\int d^n \rho_+ :=\int d\rho_1 \ldots d\rho_n\, \vartheta(\rho_1)\ldots \vartheta(\rho_n)\,.
\end{equation}

\subsection{Contribution to ${\Upsilon}^{\eta\eta} _{\go CCC}$ modulo $\Hp$}

The $\eta^2C^3$ vertex in the equations on the
zero-forms $C$  resulting from equations of \cite{more}
is \begin{equation}\label{rightside} \Upsilon^{\eta\eta}(\go,C,C,C) =-\left(\dr_x B^{ \eta \eta }_3 + [\go, B^{\eta\eta  }_3]_*    +  [ {W}^\eta_1, B^{\eta   }_2]_*
  +[ {W}^{ {\eta}\eta}_2, C]_*
+\dr_x B^{ \eta  }_2\,\right). \end{equation}
 Recall, that, being $Z$-independent, ${\Upsilon}^{\eta\eta} $  is a sum of $Z$-dependent terms
that makes its $Z$-independence implicit.

As explained in Introduction, ${\Upsilon}^{\eta\eta}$ can be decomposed into parts
with different orderings of fields $\go$ and $C$. In this paper we  consider
 \be\label{projwccc}{\Upsilon}^{\eta\eta}_{\go C C C} := \Upsilon^{\eta\eta}(\go,C,C,C)\Big|_{\go CCC}\,.
 \ee
  Since the terms from $\Hp$ do not contribute to the physical
vertex such terms can be discarded. Following  \cite{4a3} equality up to  terms from $\Hp$
referred to as weak equality
is denoted as  $\approx$ .

  We start with the following results of \cite{4a3}:
 \be\label{rightsideU}% \approx
  \widehat{\Upsilon}^{\eta\eta}_{\go CCC}
\approx {\Upsilon}^{\eta\eta}_{\go C C C}=-\Big(W_{1\, \go C}^\eta \ast B_2^{\eta\, loc}+ {W}_{2\, \go CC}^{\eta\eta}\ast C+
\dr_x B^{\eta\, loc}_2\big|_{\go CCC}+
\omega\ast {B}_3^{\eta\eta}+\dr_x {B}_3^{\eta\eta}\big|_{\go CCC}  \Big)
  \q
\ee where
 \begin{multline}\label{origW1B2}
W_{1\, \go C}^\eta \ast B_2^{\eta\, loc}\approx \frac{\eta^2}{4}\int_0^1 d\mathcal{T} \T \int_0^1 d\gs
 \int d^3\rho_+ \delta\left(1-\sum_{i=1}^3 \rho_i\right) \frac{\left(z_\gga t^{ \gga}\right)\big[z_\ga y^\ga+\gs  z_\ga t^{\ga}\big]}{(\rho_1+\rho_2)}
 \times\\
\times \exp\Big\{i\mathcal{T} z_\ga y^\ga+i(1-\gs )t^{\ga}\p_{1\ga}
-i\frac{\rho_1\gs }{\rho_1+\rho_2} t^{\ga}p_{2\ga}
+i\frac{\rho_2\gs }{\rho_1+\rho_2} t^{\ga}p_{3\ga} \\
+i\mathcal{T}z^\ga\Big(-(\rho_1+\rho_2+\gs  \rho_3)t_{\ga}-(\rho_1+\rho_2)p_{1 \ga}
+(\rho_3-\rho_1)p_{2 \ga}+(\rho_3+\rho_2)p_{3\ga}\Big) \\
+iy^\ga\Big(\gs  t_{\ga}-\frac{\rho_1}{\rho_1+\rho_2}p_{2 \ga}
+\frac{\rho_2}{\rho_1+\rho_2}p_{3\ga}\Big)\Big\}\go CCC\,,
\end{multline}

\begin{multline}\label{origW2C}
 {W}_{2\, \go CC}^{\eta\eta}\ast C\approx-\frac{\eta^2}{4}\int_0^1 d\mathcal{T}\,\T
 \int d^4\rho_+\, \delta\left(1-\sum_{i=1}^4 \rho_i\right)
 \frac{\rho_1 \left(z_\gga t^{\gga}\right)^2}{(\rho_1+\rho_2)(\rho_3+\rho_4)}\times\\
\times \exp\Big\{i\mathcal{T}z_\ga y^\ga+i\mathcal{T}z^\ga\Big((1-\rho_2)t_{\ga}
-(\rho_3+\rho_4)p_{1\ga}+(\rho_1+\rho_2)p_{2 \ga}+p_{3 \ga}\Big)+i y^\ga t_{\ga} \\
+\frac{\rho_1\rho_3}{(\rho_1+\rho_2)(\rho_3+\rho_4)}\left(i y^\ga t_{ \ga}
+it^{ \ga}p_{3\ga}\right)+i\left(\frac{(1-\rho_4)\rho_2}{\rho_1+\rho_2}
+\rho_4\right)t^{\ga}p_{1\ga}-i\frac{\rho_4\rho_1}{\rho_3+\rho_4}t^\ga p_{2\ga}\Big\} \go CC C,
\end{multline}

\begin{multline}\label{FFFFFFFFk}
\dr_x B^{\eta\, loc}_2\big|_{\go CCC}\approx \frac{\eta^2}{4}\int_0^1 d\mathcal{T}
\int_0^1 d\xi\int d^3\rho_+\,
 \delta\left(1-\sum_{i=1}^3\rho_i\right)\left(z_\ga y^\ga\right)\Big[\left(\mathcal{T}z^\ga
 -\xi y^\ga\right)t_{ \ga}\Big]\times\\
\times \exp\Big\{i\mathcal{T}z_\ga y^\ga+i(1-\rho_2)t^\ga p_{1\ga}
-i\rho_2 t^\ga p_{2\ga} +i\mathcal{T}z^\ga\Big(-(\rho_1+\rho_2)t_{\ga}
-\rho_1 p_{1 \ga}+(\rho_2+\rho_3)p_{2 \ga}+p_{3\ga}\Big) \\
+iy^\ga\Big(\xi(\rho_1+\rho_2)t_{\ga}+\xi\rho_1 p_{1 \ga}
-\xi(\rho_2+\rho_3)p_{2 \ga}+(1-\xi)p_{3\ga}\Big) \Big\}\go CCC\,,
\end{multline}

\begin{multline}\label{wB3modH+}
\omega\ast {B}_3^{\eta\eta}\approx-\frac{\eta^2}{4} \int_0^1 d\mathcal{T}\, \mathcal{T}
 \int d^3 \rho_+ \delta\left(1-\sum_{i=1}^3 \rho_i\right)   \int_0^1 d\xi\, \frac{\rho_1\,
\left[z_\ga\left(y^\ga+t^\ga\right)\right]^2   }{(\rho_1+\rho_2)(\rho_1+\rho_3)}\times\\
\times\exp\Big\{i\mathcal{T}z_\ga y^\ga
+i\mathcal{T} z^\ga\Big(-t_{ \ga}-(\rho_1+\rho_3)p_{1\ga}+(\rho_2-\rho_3)p_{2\ga}
+(\rho_1+\rho_2)p_{3\ga}\Big)+iy^\ga t_{\ga}\\
+i(1-\xi)y^\ga\left(\frac{\rho_1}{\rho_1+\rho_2}p_{1\ga}
-\frac{\rho_2}{\rho_1+\rho_2}p_{2\ga}\right)
+i\xi\, y^\ga\left(\frac{\rho_1}{\rho_1+\rho_3}p_{3\ga}-\frac{\rho_3}{\rho_1+\rho_3}p_{2\ga}\right) \\
+i\frac{(1-\xi)\rho_1}{\rho_1+\rho_2}t^{\ga}p_{1\ga}
-i\left(\frac{(1-\xi)\rho_2}{\rho_1+\rho_2}+\frac{\xi\rho_3}{\rho_1+\rho_3}\right) t^{\ga}p_{2\ga}
+i\frac{\xi\rho_1}{\rho_1+\rho_3}t^\ga p_{3\ga}\Big\} \go CCC,
\end{multline}

\begin{multline}\label{kuku5}
\dr_x {B}_3^{\eta\eta}\big|_{\go
CCC}\approx \frac{\eta^2}{4} \int_0^1 d\mathcal{T}\, \mathcal{T}
\int d^3 \rho_+ \delta\left(1-\sum_{i=1}^3 \rho_i\right) \int_0^1
d\xi\, \frac{\rho_1\, (z_\ga y^\ga)^2
  }{(\rho_1+\rho_2)(\rho_1+\rho_3)}\times\\
\times\exp\Big\{i\mathcal{T}z_\ga y^\ga+i\mathcal{T} z^\ga
\Big(-(\rho_1+\rho_3)(t_{\ga}+p_{1\ga})+(\rho_2-\rho_3)p_{2\ga}
+(\rho_1+\rho_2)p_{3\ga}\Big)+it^\ga p_{1\ga} \\
+i(1-\xi)y^\ga\left(\frac{\rho_1}{\rho_1+\rho_2}(t_{ \ga}+p_{1\ga})-\frac{\rho_2}{\rho_1+\rho_2}p_{2\ga}\right)+\xi\, y^\ga\left(\frac{\rho_1}{\rho_1+\rho_3}p_{3\ga}-\frac{\rho_3}{\rho_1+\rho_3}p_{2\ga}\right)\Big\}\go CCC.
\end{multline}
The sum of \rhss of \eq{origW1B2}-\eq{kuku5} yields  $\widehat{\Upsilon}^{\eta\eta}_{\go CCC} (Z;Y) $.

 Note, that all terms on the \rhss of \eq{origW1B2}-\eq{kuku5}  contain  no
$p_j{}_\ga p_i{}^\ga$ contractions in the exponentials, hence being spin-local
 \cite{4a3}. Thus   $\widehat{\Upsilon}^{\eta\eta}_{\go CCC} (Z;Y) $ is also spin-local.

Let us emphasize that  only  the   full expression for  $\Upsilon^{\eta\eta}_{\go CCC}(Y) $ \eq{projwccc}
is $Z$-independent, while $\widehat{\Upsilon}^{\eta\eta}_{\go CCC} (Z;Y) $  \eqref{rightsideU}
with discarded terms in $\Hp$  is not.
This does not allow one to find  manifestly
$Z$-independent expression for $ {\Upsilon}^{\eta\eta}_{\go CCC} $ by setting for instance
   $Z=0$ in Eqs.~\eq{origW1B2}-\eq{kuku5}.

   In this paper $Z$-dependence of  $\widehat{\Upsilon}^{\eta\eta}_{\go CCC}(Z;Y)$
is eliminated   modulo terms in $\Hp$   by virtue of  partial integration
  and the Schouten identity. As a result,
 $$%\Upsilon^{\eta\eta}_{\go CCC}(Y)\approx
 \widehat{\Upsilon}^{\eta\eta}_{\go CCC}(Z;Y)\approx \widehat{\widehat{\Upsilon}}  {\,}^{\eta\eta}_{\go CCC}(Y),$$
  where $\widehat{\widehat{\Upsilon}}  {\,}^{\eta\eta}_{\go CCC}(Y)$  is manifestly  spin-local  and   $Z$-independent.
   Since $\Hp_0$-terms do not contribute to the vertex by Z-dominance Lemma \cite{2a1}
% $ \Upsilon{\,}^{\eta\eta}_+=0$ and

 $$\Upsilon^{\eta\eta}_{\go CCC}(Y)=\widehat{\widehat{\Upsilon}}  {\,}^{\eta\eta}_{\go CCC}(Y)\,.$$

  Our goal is to find the manifest form of
$\widehat{\widehat{\Upsilon}}  {\,}^{\eta\eta}_{\go CCC}(Y)$.

  \section{Calculation scheme}
\label{Schema}

The calculation scheme   is as follows.

\begin{itemize}

\item I. We start from the expression Eqs.~\eq{origW1B2}-\eq{kuku5} for the vertex obtained in \cite{4a3}.

\bigskip

\item II. To $z$-linear pre-exponentials. \\
 Using partial integration and the Schouten identity
we transform Eqs.~\eq{origW1B2}-\eq{kuku5}  to the form with  $z$-linear pre-exponentials modulo
weakly $Z$-independent
%EEEEEEEEEEEEEEE
 (cohomology) terms.
These expressions are collected in Section \ref{zlinear}, Eqs.~\eq{RRwB3modH+}-\eq{W2C3gr1}.
The respective cohomology terms being a part of  the vertex
$\Upsilon^{\eta\eta}_{\go CCC} $
%etaeta
are presented in Section \ref{Main}\,.

\bigskip
\item
III. Uniformization.\\ We observe that the \rhss of Eqs.~\eq{RRwB3modH+}-\eq{W2C3gr1}
can be re-written modulo cohomology and weakly zero terms  in a form
of integrals $\int d\Gamma$ over the same   integration domain  $\II$
\begin{equation}\label{comexp}
  \int d\Gamma \, z_\ga f^\ga (y,t,p_1,p_2,p_3\vert \T,\xi_i,\rho_i)\Ee\, \go CCC\,,
\end{equation}
where the  integrand contains an overall exponential function  $\Ee$
\begin{equation}
\label{Ee}\Ee= \Ez E,
\end{equation}
\be
\label{expz}
  \Ez:=\exp i\Big\{\T z_{\ga}(y  + \Pz{})^{\ga}   \Big\}
\q\ee
 \bee
  \label{Egx=}
  &&E:=\exp i  \Big\{
 -   \gx_2  \ff{\gr_2}{(1-\gr_1-\gr_4 )(1-\gr_3 )}\,\,\big( y  + \Pz{}\big)^\ga y_{\ga}
\\ \nn &&
+  \gx_1   \ff{ \gr_2}{(1-\gr_1-\gr_4 )(1-\gr_3)}\big( y  + \Pz{}\big)^\ga\tilde{t}{}_{\ga}
\\ \nn &&+\ff{ \gr_3 }{(1-\gr_1-\gr_4 ) }\,\,  ( p_3+p_2)^{\ga}
  y_{\ga}
-\ff{ \gr_3 }{(1-\gr_1-\gr_4 )(1-\gr_3 )}\,\, \gr_1 {t}{}^{\ga} y_{\ga}
  \\ \nn &&
  +      \ff{  \gr_1 }{(1-\gr_3)}      (p_1 +p_2 )^{\ga}{t}{}_{\ga}
 +   p_3{}_{\ga}  y^{\ga} +p_1{}_\ga  {t}{}^\ga
\Big\}
\,, \eee
\bee &&\label{tildet}\tilde{t}{}=\ff{\gr_1}{\gr_1+\gr_4}{t}{}\q
\\   &&\label{Pz5}
\Pz{}= \PP   + (1-\gr_4){t}{}\,,
%=(( 1-\gr_1-\gr_4)(p{}_1 +p_2) - (1-\gr_3) (p_3+p_2)   + (1-\gr_4){t}{} )\q
\\  &&\label{PP=}
\PP =( 1-\gr_1-\gr_4)(p{}_1 +p_2) - (1-\gr_3) (p_3+p_2)\,, \eee
   the integral over $\II$  is denoted as
  \begin{equation}\label{dGamma}
  \int d\Gamma=\int_0^1 d\T\int d^3 \xi_+\,  \delta\left(1-\sum_{i=1}^3 \xi_i\right)
  \int d^4 \rho_+ \, \delta\left(1-\sum_{j=1}^4 \rho_j\right)\,.
  \end{equation}

Eqs.~\eq{RRwB3modH+}-\eq{W2C3gr1} transformed to the   form \eq{comexp}
are collected in Section \ref{uniform}, Eqs.~\eqref{F1}-\eqref{F4}.

\item IV. Elimination of $\gd$-functions.\\
Using partial integration  and the Schouten identity  we eliminate
the all factors of $\gd(\gr_i)$,
% EEEEEEEEEEEEEEE
 $\gd(\gx_{1 })$ and $\gd(\gx_{ 2})$ from Eqs.~\eqref{F1}-\eqref{F4}. The result is presented in Section \ref{Eli0},  Eqs.~\eqref{rightsideUUNI==}-\eq{FRest3}.

\item V. Final step.\\ Finally, we  show in Section \ref{proof} that a sum of
the \rhss of Eqs.~\eqref{FRest1}-\eqref{FRest3} %being  an integral   over region $\II$
 %EEEEEEEEEEEEEEEEEEEEE
  is   $Z$-independent
    up  to   $\Hp$.
\end{itemize}

By collecting all resulting $Z$-independent terms  we finally
   obtain  the manifest expression
 for vertex $\Upsilon^{\eta\eta}_{ \go CCC}$, being a sum of  expressions  \eq{go B3modHcoh}-\eq{ERRGTC}.

\section{Main result $\Upsilon^{\eta\eta}_{\go CCC}$}
\label{Main}

Here
the final manifestly $Z$-independent $\go CCC$ contribution  to the equations is presented.

Vertex $\Upsilon^{\eta\eta}_{\go CCC}$ is
 \be\label{upsrES}
\Upsilon^{\eta\eta}_{\go CCC}=\sum_{j=1}^{11} J_j\,
\ee
with $J_i$  given in Eqs.~\eq{go B3modHcoh}-\eq{ERRGTC}.
Note that the integration
regions  may differ for different terms $J_j$
in the vertex, depending on their genesis.

Firstly we note that     $B^{\eta\eta}_3$ \eqref{B3modH=1406}, that contains a $Z$-independent part,  generates cohomologies both from $\go*B^{\eta\eta}_3$  and from $\dr_x B^{\eta\eta}_3$,
\begin{equation}\label{go B3modHcoh}
J_1= - \ff{   \eta^2  }{4 }  \int d\Gamma\,  \delta(\xi_3)\ff{\gr_2}{(\gr_2+\gr_1    )(\gr_2+\gr_3)} \gd(\gr_4) E\,  \go CCC,
\end{equation}
 \begin{equation}\label{dx B3modHcoh}
J_2= \ff{   \eta^2  }{4 }      \int d\Gamma\,  \delta(\xi_3)\ff{\gr_2}{(\gr_2+\gr_4    )
(\gr_2+\gr_3)} \gd(\gr_1) E\, \go CCC\,.
\end{equation}
Recall that $E$ and $d\Gamma$ are defined in \eq{Egx=} and  \eq{dGamma}, respectively.
(Note, that, here and below, the integrands on the \rhss of expressions for $J_i$ are $\T$-independent, hence the factor of $\int_0^1 d\T$ in $d\Gamma$ equals one.)

%\bigskip

Other
cohomology terms are collected  from \eqref{FRest1}, \eqref{FRest2}, \eqref{FRest3},
\eq{lostcohomo}, \eq{lostcohomo2}, \eqref{D4}, \eqref{D6}, \eqref{D7}
and \eqref{ERRGT}, respectively,

\begin{multline}\label{Result2}
J_3=     -\ff{i   \eta^2  }{4 }  \int d\Gamma\,  \delta(\xi_3) \ff{1}{(\gr_2 +\gr_3)(1-\gr_3) }
\Big\{\gr_2 {t}{}^\ga (p_1+p_2 ){} _\ga  \big[\overrightarrow{\p}_{\gr_2}-\overrightarrow{\p}_{\gr_3}\big]    \\
  + \gr_2      ( p_1{}+ p_2)^{\ga}   ( p_3{}+p_2)_{\ga}
       \big[\overrightarrow{\p}_{\gr_4}-\overrightarrow{\p}_{\gr_1}\big] +\gr_2  {t}{}^\ga  ( p_3{}  +p_2{} )_\ga  \big[\overrightarrow{\p}_{\gr_2}-\overrightarrow{\p}_{\gr_1}\big]+\ff{ \gr_1+\gr_4}{ (1-\gr_3)  } {t}{}^\ga  (p_1+p_2 ){} _\ga
   \Big\}E\, \go C C C\,,
\end{multline}

\begin{multline}\label{Result3}
J_4=\frac{i\eta^2}{4}\int d\Gamma\, \frac{\delta(\xi_3)}{1-\rho_3}\Big(
        %[-\overrightarrow{\p}_{\gr_1}+\overrightarrow{\p}_{\gr_2}]
   -       \ff{ \gr_3}{(1-\gr_1-\gr_4 )^2(1-\gr_3)}
    {t}{}^\gga  y_\gga
  -  \ff{ \gr_2}{(1-\gr_1-\gr_4 )(1-\gr_3)}
   {t}{}^\gga  y_\gga
      [-\overrightarrow{\p}_{\gr_1}+\overrightarrow{\p}_{\gr_2}] \\
   -        \ff{ \gr_2}{(1-\gr_1-\gr_4 )(1-\gr_3)}
      ( p_1{}+ p_2)^{\gga} (y+\tilde{t}{}) _{\gga}
    [ \overrightarrow{\p}_{\gr_4}-\overrightarrow{\p}_{\gr_1}]
       \Big)E\go C C C\,,
\end{multline}

% %Cohomology term of \eq{dB3goB3rest2W1B2sub18+BPF}
\begin{multline}\label{Result4}
J_5=-i  \ff{   \eta^2  }{4 } \int d\Gamma\,\delta(\xi_3)\Big[1+ \gx_1(\overrightarrow{\p}_{\gx_1}-\overrightarrow{\p}_{\gx_2})\Big]
   \Big\{
 \ff{ -\gr_2 }{(1-\gr_1-\gr_4 )^2(1-\gr_3) ( \gr_1+\gr_4 ) }
   (p_3{}^{\ga}+p_2{}^{\ga})^\gga {t}{}_{\gga}  \\
- \ff{ \gr_3 }{(1-\gr_1-\gr_4 )^2(1-\gr_3 )^2}\,\,  {t}{}^{\ga} y_{\ga}
  +    \ff{ 1}{(\gr_2 +\gr_3)(1-\gr_3) ( \gr_1+\gr_4 ) }        (p_1 +p_2 )^{\ga}{t}{}_{\ga}
 \Big\} E\, \go C C C\,,
\end{multline}

%Cohomology term of \eq{lostcohomo}
\begin{equation}\label{Result5}
J_6=i\ff{   \eta^2  }{4 } \int d\Gamma\,\delta(\xi_3)      \ff{ \gr_2}{(1-\gr_1-\gr_4 )(1-\gr_3)^2(\gr_1+\gr_4)}
      ( p_1{}+ p_2)^{\gga} ( {t}{}) _{\gga}E \go C C C\,,
\end{equation}%Cohomology term of \eq{lostcohomo2}
\begin{multline}\label{Result6}
J_7=- \frac{\eta^2}{4}\int d\Gamma\, \delta(\xi_3)\, \xi_1
\ff{ \gr_2\gr_2}{(\gr_2 +\gr_3)^3(1-\gr_3)^3( \gr_1+\gr_4 ) }\times\\
\times \big( y+  (1-\gr_1-\gr_4 )( p_1{}  +p_2{} )+  (1 -\gr_4 ){t}{}  \big)^\gga
\big( y +   \tilde{t}{}  \big)_{\gga}
{t}{}^{\ga} y_\ga       E \, \go CCC\,,
\end{multline}
\begin{equation}\label{Result1}
J_8=-   \ff{   \eta^2  }{4 }\int d\Gamma \, \delta(\rho_3)
 \Big(       \gr_1\gd(\gx_3 )
 + \Big[  i {\gd(\gr_4)} -( p_2{}_\ga+ p_1{}_\ga) {t}{}^{\ga}\Big]
  \Big\{  i        \gd(\gx_3 )+
   \tilde{t}{}^{\gga} y_\gga\Big\}\Big)E\, \go CCC\,,
\end{equation}

\begin{multline}\label{goB3modH1406gr1C}
J_9= i\eta^2\chalf \int d\Gamma\, \delta(\rho_1)\delta(\rho_4)\delta(\xi_3)\exp \Big\{  -i\gx_2  (     p_1+p_2+{t} -  \gr_2  (p_3+p_2))_{\ga} (y )^{\ga} \\
-\gx_1 ( y+    p_1+p_2 -  \gr_2  (p_3+p_2))_\gga ( {t})^\gga
 + ( 1-\gr_2)  (p_3+p_2)  {}^\gga y_\gga
 + p_3{}_\gga y^\gga +{t}{}^\gb p_1{}_\gb      \Big\}\go CCC\,,
\end{multline}

\begin{multline}\label{dxB3modH1406gr1C}
J_{10}=-i\eta^2\chalf \int d\Gamma\, \delta(\rho_4)   \gd(\gx_1 )\gd(\gr_1)
 \,    \exp i\Big\{-\gx_2 ( y+    p_1+p_2+{t} -  \gr_2  (p_3+p_2))_{\ga} (y )^{\ga} \\
+ ( 1-\gr_2)  (p_3+p_2)  {}^\gga y_\gga  +p_3{}_\gga y^\gga +{t}{}^\gb p_1{}_\gb \Big\}\go CCC\,,
\end{multline}

\begin{multline}\label{ERRGTC}
J_{11}=\frac{i \eta^2}{4}\int d\Gamma \, \delta(\rho_1)\delta(\rho_4) y^\ga   {t} {}_\ga \exp i\Big\{ (y+\PP_0     +{t}){}^\gga ( \gx_1 {t}-  \gx_2 y)_\gga       + ( 1-\gr_2)  (p_3+p_2)  {}^\gga y_\gga
  \\
 + p_3{}_\gga y^\gga +{t}{}^\gb p_1{}_\gb \Big\}\go CCC\,.
\end{multline}

Let us emphasize, that neither exponential function $E$ \eq{Egx=}
nor the exponentials on the \rhss of Eqs.~\eq{goB3modH1406gr1C}-\eq{ERRGTC}
contain $\p_i{}_\ga  \p_k{}^\ga$ terms.
Hence, as anticipated, all  $J_j$ are spin-local.

 One can see that though having poles in pre-exponentials  these expressions  are well defined.
\\For instance   a potentially dangerous  factor  on the \rhs of \eq{go B3modHcoh}
  is dominated by 1 as follows from  the inequality
$ {\gr_2}-(\gr_1+\gr_2    )
(\gr_2+\gr_3) =-\gr_3\gr_1\le  0$\, that holds   due to the factor
of  $\prod\vartheta(\gr_i)\gd(1-\sum\gr_i )\gd(\gr_4)$.
Analogous simple reasoning  applies to the \rhs of \eq{dx B3modHcoh}.

The case of  \eq{Result2}-\eq{Result6} is a  bit more tricky.
By partial integration   one obtains from \eq{Result2}-\eq{Result4}
\bee\label{RRult4+} &&
 J_3+J_4+J_5 =  \ff{i   \eta^2  }{4 }  \int d\Gamma\,
\delta(\xi_3)\ff{1 }{(\gr_2 +\gr_3)(1-\gr_3) }\Big\{
-
\gd({\gr_3})  {t}{}^\ga (p_1+p_2 ){} _\ga
\\&&\nn
  +  [ \gd({\gr_4})-\gd({\gr_1})]\gr_2
  ( p_1{}+ p_2)^{\ga}   ( p_3{}+p_2)_{\ga}
  +  {t}{}^\ga  ( p_3{}  +p_2{} )_\ga      -\gd({\gr_1})\gr_2  {t}{}^\ga  ( p_3{}  +p_2{} )_\ga
       \,
\\&&\nn
    -\gd({\gr_1})  \ff{ \gr_2}{ (1-\gr_3)}
   {t}{}^\gga  y_\gga
   + [ \gd({\gr_4})-\gd({\gr_1})]       \ff{ \gr_2}{ (1-\gr_3)}
      ( p_1{}+ p_2)^{\gga} (y+\tilde{t}{}) _{\gga}
  \\ &&\nn
 -    \gd({\gx_2})
 %\Big[  \gx_1\gd(\gx_2)\Big]
   \Big(
 \ff{ -\gr_2 }{(\gr_2 +\gr_3)( \gr_1+\gr_4 ) }
   (p_3{}^{\ga}+p_2{}^{\ga})^\gga {t}{}_{\gga}
   \\&&\nn
- \ff{ \gr_3 }{(\gr_2 +\gr_3) (1-\gr_3 ) }\,\,  {t}{}^{\ga} y_{\ga}
  +    \ff{ 1}{  ( \gr_1+\gr_4 ) }        (p_1 +p_2 )^{\ga}{t}{}_{\ga}
\Big) \Big\} E\, \go C C C\,.
\eee
Using that, due to the  factor of $\gd(1-\sum \gr_i)$,
for positive $\gr_i $ it holds
\bee&&\label{nopoles}
 \ff{\gr_2}{(\gr_3+\gr_2)(1-\gr_3)}-1 =-\ff{\gr_3(1-(\gr_3+\gr_2    ))}{(\gr_3+\gr_2)(1-\gr_3)}\le  0
 \q\\ \label{nopoles3} && \ff{ 1}{(\gr_2 +\gr_3)(1-\gr_3)  }\,\le
\ff{ 1}{(\gr_2 +\gr_3)(1-\gr_3-\gr_2)  }=
\ff{ 1}{ ( \gr_3+\gr_2)  }+\ff{ 1}{  ( \gr_1+\gr_4 )  }\,,
\eee
one can make sure that each of the expressions with poles in the pre-exponential   in Eqs.~\eq{Result5}, \eq{Result6} and
 \eq{RRult4+}
can be represented in the form of a sum of integrals with integrable pre-exponentials.
For instance, the potentially dangerous
factor in \eq{Result6},  by virtue of \eq{nopoles} and \eq{nopoles3} satisfies
\be    \ff{ \gr_2\gr_2}{(\gr_2 +\gr_3)^3(1-\gr_3)^3( \gr_1+\gr_4 ) }\le
  \ff{ 1}{ (1-\gr_3)( \gr_1+\gr_4 ) }+
 \ff{1}{(\gr_3+\gr_2) }
 +\ff{ 1}{  ( \gr_1+\gr_4 )  }\,.\quad\label{xx}
\ee
 Each of the   terms on the \rhs of Eq.~(\ref{xx}) is integrable, because integration
  is  over a three-dimensional compact area $\sum\gr_i=1$ in the positive quadrant.
For instance consider the first term. Swopping   $\gr_4\leftrightarrow\gr_2$ one has
  \bee
\int d^4 \gr_+ \gd(1-\sum_1^4 \gr_i)\ff{1}{(1-\gr_3 ) ( \gr_1+\gr_2)}=
\int d^3 \gr_+ \vartheta(1-\sum_1^3 \gr_i)\ff{1}{(1-\gr_3 ) ( \gr_1+\gr_2)}=\\ \nn
-\int_0^1 d \gr_1 \int_0^{1-\gr_1} d \gr_2
  \ff{\log( \gr_1+\gr_2)}{   ( \gr_1+\gr_2)}=\half\int_0^1 d \gr_1   \log^2( \gr_1 )\,,
     \eee
 which is integrable.

 Analogously
other  seemingly dangerous  factors can be shown to be harmless as well.

\section{To $z$-linear pre-exponentials}
 \label{zlinear}
Step II of the calculation scheme of Section \ref{Schema}  is to transform  \rhss of  Eqs.~\eq{origW1B2}-\eq{kuku5} to   $Z$-independent   terms plus   terms with linear in $z$ pre-exponentials
%EEEEEEEEEEEEEEEEEEEEEEEE
(modulo $H^+$).

To this end, from  \eq{B3modH=1406}
one straightforwardly  obtains  that
 \begin{multline} \label{RRwB3modH+}
\omega \ast {B}_3^{\eta\eta}\approx J_1+ \frac{\eta^2}{4}\int d\Gamma
\frac{\gd(\xi_3)\gd(\gr_4)}{(1-\gr_1)(1-\gr_3)}\Bigg[-\gr_2 (z_\ga (y^\ga+t^\ga))(p_{1\gb}+p_{2\gb})(p_2 {}^\gb+p_3 {}^\gb) \\
+i\Big[\Big(\gd(\gr_1)+\gd(\gr_3)\Big)(1-\gr_1)(1-\gr_3)-\gd(\xi_2)\Big]
 z_\ga\Big((1-\gr_1)(p_1 {}^\ga+p_2 {}^\ga)-(1-\gr_3)(p_2 {}^\ga+p_3 {}^\ga)\Big) \\
+iz_\ga (p_1 {}^\ga+p_2 {}^\ga)(1-\gr_1)\Big(\gd(\xi_2)-\gd(\xi_1)\Big)\Bigg]
\exp\Big\{i\T z_\ga\big(y^\ga+t^\ga+(1-\gr_1)(p_1 {}^\ga+p_2 {}^\ga)
-(1-\gr_3)(p_2 {}^\ga+p_3 {}^\ga)\big) \\
+\frac{i(1-\xi_1) \gr_2}{\gr_1+\gr_2}(y^\ga+t^\ga) (p_{1\ga}+p_{2\ga})
+\frac{i\xi_1 \gr_2}{\gr_2+\gr_3}(y^\ga+t^\ga) (p_{2\ga}+p_{3\ga})-i(y^\ga+t^\ga) p_{2\ga}\Big\}\go CCC\q
\end{multline}
where  $J_1$ is the cohomology term  \eq{go B3modHcoh}.
Analogously, \begin{multline} \label{RRdxB3modH+}
\dr_x {B}_3^{\eta\eta} \approx J_2-\frac{\eta^2}{4}\int d\Gamma
\frac{\gd(\xi_3)\gd(\gr_4)}{(1-\gr_1)(1-\gr_3)}
\Bigg[-\gr_2 (z_\ga y^\ga)(p_{1\gb}+t_\gb+p_{2\gb})(p_2 {}^\gb+p_3 {}^\gb) \\
+i\Big[\Big(\gd(\gr_1)+\gd(\gr_3)\Big)(1-\gr_1)(1-\gr_3)-\gd(\xi_2)\Big]
z_\ga\Big((1-\gr_1)(p_1 {}^\ga+t^\ga+p_2 {}^\ga)-(1-\gr_3)(p_2 {}^\ga+p_3 {}^\ga)\Big) \\
+iz_\ga (p_1 {}^\ga+t^\ga+p_2 {}^\ga)(1-\gr_1)\Big(\gd(\xi_2)-\gd(\xi_1)\Big)\Bigg]
\exp\Big\{i\T z_\ga\big(y^\ga+(1-\gr_1)(p_1 {}^\ga+t^\ga+p_2 {}^\ga)-(1-\gr_3)(p_2 {}^\ga+p_3 {}^\ga)\big)
\\
+\frac{i(1-\xi_1) \gr_2}{\gr_1+\gr_2}y^\ga (p_{1\ga}+t_\ga+p_{2\ga})
+\frac{i\xi_1 \gr_2}{\gr_2+\gr_3}y^\ga (p_{2\ga}+p_{3\ga})-iy^\ga p_{2\ga}+it^\gb p_{1\gb}\Big\}\go CCC
\end{multline}
with  $J_2$ \eq{dx B3modHcoh}.

Using the Schouten identity and partial integration   one obtains   from Eqs.~\eq{origW1B2}-\eq{FFFFFFFFk}, respectively,
 \begin{multline}\label{RW1B2BP=}
W_{1 \, \go C}^\eta \ast B_2^\eta\approx \frac{\eta^2}{4}\int_0^1 d\T
\int_0^1 d\tau\int_0^1 d\gs_1 \int_0^1 d\gs_2\Bigg[i(z_\ga t^\ga)\gd(1-\tau) \\
+\frac{z_\ga(p_2 {}^\ga+p_3 {}^\ga)}{1-\gt}\Big(i\big(\gd(\gs_1)-\gd(1-\gs_1)\big)
-\big[y^\ga+p_1 {}^\ga +p_2 {}^\ga-\gs_2(p_2{}^\ga+p_3 {}^\ga)\big]t_\ga\Big)\Bigg]\exp\Big\{i\T z_\ga y^\ga \\
+i\T z_\ga\Big(\tau(p_1 {}^\ga +p_2 {}^\ga)-((1-\tau)+\gs_2\tau)(p_2 {}^\ga +p_3 {}^\ga)
+\big(\gs_1+\tau(1-\gs_1)\big)t^\ga\Big)+it^\ga p_{1\ga} \\
+i\gs_1\big[y^\ga+p_1 {}^\ga +p_2 {}^\ga-\gs_2(p_2{}^\ga+p_3 {}^\ga)\big]t_\ga
-i\Big(\gs_2 p_3 {}^\ga-(1-\gs_2)p_2 {}^\ga\Big)y_\ga\Big\}\go CCC\,,
\end{multline}
\begin{multline}\label{W2C3gr1}
W_{2\, \go CC}^{\eta\eta}\ast C\approx -\frac{i\eta^2}{4}
\int d\Gamma\, \gd(\xi_3)\gd(\gr_3)\frac{(z_\gga t^\gga)}{\gr_1+\gr_4}
\Big[-\gr_1\big( \gd(\gr_4)+i t^\ga(p_{1\ga}+p_{2\ga})\big)+\xi_1\gd(\xi_2)\Big]\times\\
\times \exp\Big\{i\T z_\ga y^\ga+i\T z_\ga
\Big((1-\gr_1-\gr_4)(p_1 {}^\ga+p_2 {}^\ga)-(1-\gr_3)(p_2{}^\ga+p_3 {}^\ga)+(1-\gr_4)t^\ga\Big) \\
+iy^\ga\left(\frac{\xi_1 \gr_1}{1-\gr_2}t_\ga+p_{3\ga}\right)
+i\left(1-\gr_1-\frac{\xi_1 \gr_1\gr_2}{1-\gr_2}\right)t^\ga p_{1\ga}-i(1-\xi_1)\gr_1 t^\ga p_{2\ga}
+i\frac{\xi_1 \gr_1}{1-\gr_2}t^\ga p_{3\ga} \Big\}\go CCC\,,
\end{multline}
\begin{multline}\label{FFFFFFFFk=}
\dr_x B_2^\eta\approx\frac{i\eta^2}{4} \int d\Gamma\, \gd(\xi_3)\gd(\gr_4)\, (z_\ga y^\ga)
\Big[it^\gga(p_{1\gga}+p_{2\gga})+\gd(\gr_4)- \gd(\gr_1) \Big]\times\\
\times \exp\Big\{i\T z_\ga y^\ga
+i\T z_\ga\big((1-\gr_1-\gr_4)(p_1 {}^\ga+ p_2 {}^\ga)-(1-\gr_3)(p_2 {}^\ga +p_3 {}^\ga)+(1-\gr_4)t^\ga\big) \\
+i(1-\gr_2)t^\gb p_{1\gb}-i\gr_2 t^\gb p_{2\gb}
+i\xi_2 y^\ga \Big((\gr_1+\gr_2)t_\ga+\gr_2 p_{1\ga}-(1-\gr_2)p_{2\ga}-p_{3\ga}\Big)
+iy^\ga p_{3\ga} \Big\}\go CCC.
\end{multline}

\section{Generalised Triangle identity}
\label{SecGTid}

 Here   a  useful identity  playing the key role in our computations is introduced.

   For  any $F(x,y)$
consider
\bee\label{GTH+F}
 &&I= \int_{[0,1]}   {d
\gt\,}\int     d^3
\gx_+
   \gd(1-\gx_1-\gx_2-\gx_3 )   \\ \nn&&
 z^\gga \Big[  (a_2-a_1)_\gga \gd(\gx_3)+   (a_3-a_2)_\gga \gd(\gx_1)
+   (a_1-a_3)_\gga \gd(\gx_2)\Big] F \big(
 \gt  z_\gb  P^\gb\,,   ( -\gx_1 a_1-\gx_2 a_2-\gx_3 a_3)_\ga  P^\ga \big)\,
\eee
  with arbitrary         $\gt, \gx$- independent $P$ and   $a_i$.

Let $G(x,y)$ be a solution to differential equation
\be\label{difvim}
\ff{\p}{\p x} G(x,y)=  \ff{\p}{\p y}F (x,y)\,. \ee
%for instance,\be\label{difvim=}G= \int_{a}^{x} d t \ff{\p}{\p y}(t,y)\,.\ee
Hence
\bee\label{GTHF0}
&& I   =   \int_{[0,1]}   {d
\gt\,}\int     d^3
\gx_+ \gd(1-\gx_1-\gx_2-\gx_3 )   \\ \nn&&
 (a_1-a_3)^\ga(a_3-a_2)_\ga
 \overrightarrow{\p}_\gt G   \big(
 \gt  z_\gb P^\gb \,, (-\gx_1 a_1-\gx_2 a_2-\gx_3 a_3)_\ga  P^\ga \big). \eee
Note that  there is a factor of $(a_1-a_3)^\ga(a_3-a_2)_\ga$     equal     to the  area
of triangle spanned
by the vectors $a_1\,,a_2\,, a_3$ on the \rhs of \eq{GTHF0}.

This identity is closely related to identity (3.24) of \cite{4a1}, that, in turn,   expresses
{\it triangle identity} of \cite{Vasiliev:1989xz}.
Hence,   \eq{GTHF0} will be referred to as
{\it Generalised Triangle identity} or   {\it GT identity}.

Note that,
  for appropriate $G$   partial integration on the \rhs of \eq{GTHF0}
 in $\gt$  gives $z$-independent (cohomology) term  plus $\mathcal{H} ^+$-term. Namely,
\bee\label{GTHF0pi}
&& I   =  -   \int     d^3
\gx_+   \gd(1-\gx_1-\gx_2-\gx_3 )   \\ \nn&&
 (a_1-a_3)^\ga(a_3-a_2)_\ga
   G   \big(
  0\,, (-\gx_1 a_1-\gx_2 a_2-\gx_3 a_3)_\ga  P^\ga \big)
 \\ \nn&&+ \int     d^3_+
\gx   \gd(1-\gx_1-\gx_2-\gx_3 )   \\ \nn&&
 (a_1-a_3)^\ga(a_3-a_2)_\ga
   G   \big(
     z_\gb P^\gb \,, (-\gx_1 a_1-\gx_2 a_2-\gx_3 a_3)_\ga  P^\ga \big)
  . \eee
The second term on the \rhs belongs to $\Hp$       if $G$ is of the form \eq{class} satisfying \eq{limit}.

To prove GT identity let us perform
  partial integration on the \rhs of \eq{GTH+F} with respect to $\gx_i$. This yields
\bee\label{GTH+F=}
 && I= \int_{[0,1]}   {d
\gt\,}\int    {d^3
\gx_+\,}  \gd(1-\gx_1-\gx_2-\gx_3 )  \\ \nn&&
 \Big[
 z^\gga  (a_3-a_2)_\gga P^\ga a_1{}_\ga
+z^\gga  (a_1-a_3){}_\gga P^\ga  a_2{}_\ga
+z^\gga  (a_2-a_1){}_\gga P^\ga a_3{}_\ga
\Big]\times\\ \nn&&  \ff{\p}{\p y} F \big(
 \gt  z_\ga  P^\ga \,,\,\,-(\gx_1 a_1+\gx_2 a_2+\gx_3 a_3)_\ga  P^\ga \big)\,. \eee
The Schouten identity yields
\bee
 \Big[
z^\gga  a_1{}_\gga P^\ga(a_3-a_2)_\ga
+z^\gga  a_2{}_\gga P^\ga(a_1-a_3)_\ga
+z^\gga  a_3{}_\gga P^\ga(a_2-a_1)_\ga
\Big]
=\\ \nn
\Big[z^\gga  P_\gga \big\{
 a_1{}^\ga(a_3-a_2)_\ga
+  a_2^\ga(a_1-a_3)_\ga
+  a_3^\ga(a_2-a_1)_\ga\big\}
\\ \nn
+z^\gga  (a_3-a_2)_\gga P^\ga a_1{}_\ga
+z^\gga  (a_1-a_3){}_\gga P^\ga  a_2{}_\ga
+z^\gga  (a_2-a_1){}_\gga P^\ga a_3{}_\ga
\Big].
\eee
One can observe  that
\bee \Big[z^\gga  (a_3-a_2)_\gga P^\ga a_1{}_\ga
+z^\gga  (a_1-a_3){}_\gga P^\ga  a_2{}_\ga
+z^\gga  (a_2-a_1){}_\gga P^\ga a_3{}_\ga
\Big]=\\ \nn
- \Big[
z^\gga  a_1{}_\gga P^\ga(a_3-a_2)_\ga
+z^\gga  a_2{}_\gga P^\ga(a_1-a_3)_\ga
+z^\gga  a_3{}_\gga P^\ga(a_2-a_1)_\ga
\Big]\q
\eee
 whence it follows \eq{GTHF0}.

A useful   particular case of GT identity is that with $F(x,y) =f(x+y)$, namely
 \bee\label{GTH+==0}
 &&  \int_{[0,1]}   {d
\gt\,}\int     {d^3
\gx_+\,}  \gd(1-\gx_1-\gx_2-\gx_3 )    z^\gga \Big[  (a_2-a_1)_\gga \gd(\gx_3) \\ \nn&&
+   (a_3-a_2)_\gga \gd(\gx_1)
+   (a_1-a_3)_\gga \gd(\gx_2)\Big]  f\big(
 (\gt  z-\gx_1 a_1-\gx_2 a_2-\gx_3 a_3)_\ga  P^\ga \big) \quad\\ \nn
&&     =   -  \int_{[0,1]}   {d \gt\,}\int   {d^3
\gx_+\,}  \gd(1-\gx_1-\gx_2-\gx_3 )   \\ \nn&&
  (a_1-a_3)^\ga(a_3-a_2)_\ga
 \overrightarrow{\p}_\gt f \big(
 (\gt  z-\gx_1 a_1-\gx_2 a_2-\gx_3 a_3)_\ga  P^\ga \big) \,. \eee

 \section{Uniformization}
\label{uniform}

Step III of Section   \ref{Schema} is to uniformize the \rhs's of
 Eqs.~\eq{RRwB3modH+}-\eq{FFFFFFFFk=} putting them into  the  form \eq{comexp}, where GT identity  \eq{GTH+F} plays an important role.
Details of  uniformization are given in Appendix B
 (p. \pageref{Auniform}).

   As a result, Eq.~\eq{rightsideU} yields
 \begin{equation} \label{rightsideUUNI=}
\widehat{\Upsilon}^{\eta\eta}_{\go CCC}\Big|_{\text{mod}\, cohomology}\approx
\sum_{j=1}^4  F_j
\end{equation}
with $F_j$ presented  in \eq{F1}-\eq{F4}.
%HERE

 Note that different terms of $F_j$ will be considered separately in what is follows.
 For the future convenience   the  underbraced terms are re-numerated,
 being denoted as $F_{j,k}$, where $j$ refers to $F_j$ while $k$ refers to the
 respective underbraced term in the expression for $F_j$.
For instance, $F_1=F_{1,1}+F_{1,2}+F_{1,3}+F_{1,4}$, {\it{etc}}.

\begin{multline}\label{F1}
-\go\ast B_3^{\eta\eta}\Big|_{mod\, \delta(\rho_1)\&\delta(\T)}\approx F_1 :=-\frac{\eta^2}{4}\int d\Gamma\, \frac{\delta(\xi_3)\delta(\rho_4)}{(1-\rho_1-\rho_4)(1-\rho_3)}\Big[\underbrace{\rho_2 (z_\beta \PP^\beta)(p_{1\ga}+p_{2\ga})(p_2 {}^\ga+p_3 {}^\ga)}_1 \\
+\underbrace{ i\delta(\rho_3)(1-\rho_1-\rho_4)(1-\rho_3) (z_\ga \PP^\ga)}_2 +\underbrace{-i\xi_1\delta(\xi_2)(z_\ga \PP^\ga)}_3 \\
+\underbrace{i(1-\rho_1-\rho_4)z_\ga(p_1 {}^\ga+p_2 {}^\ga)\Big(\delta(\xi_2)-\delta(\xi_1)\Big)}_4\Big]
\mathcal{E}\go CCC
\,,\end{multline}

\begin{multline}\label{F2}
-\dr_x B^{\eta\eta}_3\Big| _{mod\, \delta(\rho_1)\&\delta(\T)}\approx F_2 :=+\frac{\eta^2}{4}\int d\Gamma\, \frac{\delta(\xi_3)\delta(\rho_1)}{(1-\rho_1-\rho_4)(1-\rho_3)}\Big[\underbrace{\rho_2 (z_\beta \PP^\beta)(p_{1\ga}+p_{2\ga})(p_2 {}^\ga+p_3 {}^\ga)}_1 \\
+\underbrace{ \rho_2(1-\rho_4) (z_\beta t^\beta)t_\ga(p_2 {}^\ga+p_3 {}^\ga)}_2 +\underbrace{ \rho_2(1-\rho_4) (z_\beta t^\beta)(p_{1\ga}+p_{2\ga})(p_2 {}^\ga+p_3 {}^\ga)}_3+\underbrace{\rho_2 (z_\beta \PP^\beta)t_\ga(p_2 {}^\ga+p_3 {}^\ga)}_4 \\
+\underbrace{ i\delta(\rho_3)(1-\rho_1-\rho_4)(1-\rho_3)(z_\ga \Pz^\ga)}_5+
 \underbrace{-i\xi_1\delta(\xi_2)(z_\ga \PP^\ga)}_6+\underbrace{-i\xi_1\delta(\xi_2)(1-\rho_4)(z_\ga t^\ga)}_7 \\
+\underbrace{ i(1-\rho_1-\rho_4)z_\ga(p_1 {}^\ga+p_2 {}^\ga)
\Big(\delta(\xi_2)-\delta(\xi_1)\Big)}_8 +
\underbrace{ i(1-\rho_1-\rho_4)z_\ga t^\ga\Big(\delta(\xi_2)-\delta(\xi_1)\Big)}_9\Big]\mathcal{E}\go CCC
\,,\end{multline}

\begin{multline}\label{F3}
-\dr_xB_2^\eta-W_{2\, \go CC}^{\eta\eta}\ast C\Big|_{mod\, \delta(\T)} \approx
F_3:=-\frac{\eta^2}{4}\int d\Gamma\delta(\rho_3)\delta(\xi_3)\Bigg[
\underbrace{ i\delta(\rho_1)(z_\ga \Pz^\ga)}_1+
\underbrace{-\frac{i(z_\ga t^\ga)\, \xi_1\delta(\xi_2)}{\rho_1+\rho_4}}_2 \\
+\underbrace{ t^\ga(p_{1\ga}+p_{2\ga})z_\gga\PP^\gga}_3
+\underbrace{ i\delta(\rho_4)z_\ga (-\PP^\ga)}_4 +
\underbrace{ t^\gga(p_{1\gga}+p_{2\gga})z_\ga t^\ga\left((1-\rho_4)
-\frac{\rho_1}{\rho_1+\rho_4}\right)}_5\Bigg]\mathcal{E}\, \go CCC
\,,\end{multline}
% tut + J_8

\begin{multline}\label{F4}
-(\dr_x B_3^{\eta\eta}+\go \ast B_3^{\eta\eta})\Big|_{\delta(\rho_1)}
\Big|_{mod\, \delta(\T)}-W_{1\, \go C}^{\eta}\ast B_2^{\eta\, loc}\approx
F_4:=-\frac{\eta^2}{4}\int d\Gamma\, \frac{\delta(\xi_3)\delta(\xi_2)\, z_\ga(p_2 {}^\ga+p_3 {}^\ga)}
{(\rho_2+\rho_3)(\rho_1+\rho_4)}\times\\
\times \left(\underbrace{i\Big(\delta(\rho_1)-\delta(\rho_4)\Big)\Ee}_1+
\underbrace{ i\Ez\left(\frac{\p}{\p \rho_1}-\frac{\p}{\p \rho_4}\right)E}_2\right) \go CCC.
\end{multline}

% tut + J_9+J_{10}+J_{11}
 Note that
\be
F_{1,2}+F_{3,4}=0,
\ee
\be
F_{2,5}+F_{3,1}=0.
\ee

Let us emphasise that, by virtue \eq{EEgx14=},   each  $F_j$ is of the  form \eq{comexp} as expected.

Note that during uniformizing procedure the  vertices %$J_8$, $J_9$, $J_{10}$ and $J_{11}$
\eq{Result1} -\eq{ERRGTC} are obtained in Appendix B (p. \pageref{Auniform}).

 \section{Eliminating  $\gd(\gr_j)$ and $\gd(\gx_j)$. Result}
\label{Eli0}

The fourth step of Section \ref{Schema}   is to eliminate all
$\delta(\rho_i)$\,,
%EEEEEEEEEEEEEEE
$\delta(\xi_1)$ and $\delta(\xi_2)$ from the pre-exponentials on the \rhss
of Eqs.~\eq{F1}-\eq{F4}.

More precisely,   using partial  integration, the Schouten identity and
{  Generalised Triangle identity} \eq{GTHF0},  taking into account  Eqs.~\eq{tildet}-\eq{PP=} one finds
  that Eq.~\eq{rightsideUUNI=} yields
\begin{equation} \label{rightsideUUNI==}
\big(\widehat{\Upsilon}^{\eta\eta}_{\go CCC} - G_1-G_2-G_3\big)\big|_{\ls\mod  cohomology }\approx 0
  \q
\end{equation}
where
\begin{multline}\label{FRest1}
G_1   := J_3+\frac{\eta^2}{4}
\int d\Gamma\, \delta(\xi_3)  z_\gga\Bigg\{
  (y^\gga+\widetilde{t}^\gga) \frac{\rho_2\, t^\ga (p_{1\ga}+p_{2\ga})}{(1-\rho_1-\rho_4)(1-\rho_3)}\Ez\Bigg[\frac{\p}{\p \rho_2}-\frac{\p}{\p \rho_3}\Bigg]E \\
+  (y^\gga+\widetilde{t}^\gga) \frac{\rho_2\, (p_1 {}^\ga+p_2 {}^\ga)(p_{2\ga}+p_{3\ga})}{(1-\rho_1-\rho_4)(1-\rho_3)}\Ez \Bigg[\frac{\p}{\p \rho_4}-\frac{\p}{\p \rho_1}\Bigg]E \\
+ (y^\gga+\tilde{t}^\gga)
\frac{\rho_2\, t^\ga(p_{2\ga}+p_{3\ga})}{(1-\rho_1-\rho_4)(1-\rho_3)}
\Ez\Bigg[\frac{\p}{\p \rho_2}-\frac{\p}{\p \rho_1}\Bigg]E
+ (y^\gga+\tilde{t}^\gga)
\frac{(\rho_1+\rho_4) t^\ga (p_{1\ga}+p_{2\ga})}{(1-\rho_1-\rho_4)(1-\rho_3)}\Ee \\
+  (y^\gga+\tilde{t}^\gga)\frac{\rho_3\, t^\ga (p_{2\ga}+p_{3\ga})}{(1-\rho_1-\rho_4)^2 (1-\rho_3)}
\Ee
+\frac{\rho_2\,   t^\gga  (p_2 {}^\ga+p_3 {}^\ga)(p_{1\ga}+p_{2\ga}
+t_\ga-\tilde{t}_\ga)}{(1-\rho_1-\rho_4)(1-\rho_3)(\rho_1+\rho_4)}\Ee\Bigg\}\go CCC
\q\end{multline}
  \begin{multline}\label{FRest2}
G_2  :=  J_4
  +\frac{\eta^2}{4}\int d\Gamma\, \frac{\delta(\xi_3)}{1-\rho_3}\,z^\ga
  \Bigg\{ \frac{\rho_3 (y_\ga+\tilde{t}_\ga)t^\gga(y_\gga+\Pz_\gga) }{(1-\rho_1-\rho_4)^2(1-\rho_3)}
      \Ee \\
-\frac{\rho_2\rho_4\,  t_\ga (y^\gga+\Pz^\gga)t_\gga }
{(1-\rho_1-\rho_4)(1-\rho_3)(\rho_1+\rho_4)^2}\Ee-\frac{\rho_2\,
 (y_\ga+\tilde{t}_\ga) t^\gga(p_{1\gga}+p_{2\gga})  }{(1-\rho_1-\rho_4)(1-\rho_3)}\Ee \\
-\frac{\rho_2\,  (p_1 {}_\ga +p_2 {}_\ga )(y^\gga+\Pz^\gga)t_\gga}{(1-\rho_1-\rho_4)(\rho_1+\rho_4)(1-\rho_3)}
 \Ee
+\Ez\frac{\rho_2\, t^\gga (y_\gga+\Pz_\gga)(y_\ga+\tilde{t}_\ga)
   }
{(1-\rho_1-\rho_4)(1-\rho_3)}\Bigg[ \frac{\p}{\p \rho_1}-\frac{\p}{\p \rho_2}\Bigg]E \\
+\Ez \frac{\rho_2\, (y_\ga+\tilde{t}_\ga) (p_1 {}^\gga+p_2 {}^\gga)(y_\gga+\Pz_\gga)
   }{(1-\rho_1-\rho_4)(1-\rho_3)}\Bigg[\frac{\p}{\p \rho_1}
-\frac{\p}{\p \rho_4}\Bigg]E\Bigg\}\go CCC\q
\end{multline}
  \begin{multline}\label{FRest3}
G_{ 3}%\Big|_{\ls\mod cohomology}
 :=  J_5  +  \frac{\eta^2}{4}
\int d\Gamma\, \delta(\xi_3) \Bigg(1+\xi_1\Bigg[\frac{\p}{\p \xi_1}
-\frac{\p}{\p \xi_2}\Bigg]\Bigg)\times\\\times
 z_\ga
\Bigg\{\frac{\rho_2\,   t^\ga(p_2 {}^\gga+p_3 {}^\gga)(y_\gga+\tilde{t}_\gga)}
{(1-\rho_1-\rho_4)^2 (1-\rho_3)(\rho_1+\rho_4)}
 + \frac{-\rho_2\, t^\ga (\tilde{t}^\gga+y^\gga)(y_\gga+\Pz_\gga)
 }{(1-\rho_1-\rho_4)^2 (1-\rho_3)^2 (\rho_1+\rho_4)}
\\+\frac{-\rho_3\,  (y^\ga+\tilde{t}^\ga) (t^\gga y_\gga)}
{(1-\rho_1-\rho_4)^2 (1-\rho_3)^2}+\frac{  (y^\ga+\tilde{t}^\ga)
(p_1 {}^\gga+p_2 {}^\gga)t_\gga}{(1-\rho_1-\rho_4)(1-\rho_3)^2}  \Bigg\}\Ee \, \go CCC \q
\end{multline}
with  $J_3$,    $J_4$  and $J_5$  being the cohomology terms \eq{Result2}, \eq{Result3} and \eq{Result4}, respectively.
(Details of the derivation are presented in Appendix C (p.\pageref{AppD}).)

Note that  schematically
 \begin{equation}\label{NoDistrib}
 G_1+G_2+G_3   = \int d\Gamma\, \delta(\xi_3)
 z_\ga g ^\ga(y,t,p_1,p_2,p_3\vert \rho ,\xi ) \Ee \, \go CCC\,+  J_3+J_4+ J_5\q
\end{equation}
 as expected . Let us stress that      $g^\ga(y,t,p_1,p_2,p_3\vert \rho ,\xi)$  on the \rhs of \eq{NoDistrib} is
  free from a distributional behaviour.

\section{Final step of calculation}
\label{proof}

  Here this is shown that the sum of
  the \rhss of Eqs.~\eqref{FRest1}-\eqref{FRest3} gives a $Z$-independent
  cohomology term up to terms in  $\Hp$.

 More in detail, the   expression $ G_1+G_2+G_3   $
 of the form  \eq{NoDistrib} consists of two types of
terms with the pre-exponential  of  degree four and six in $z, y,t,p_1,p_2,p_3$, respectively.
That with degree-four  pre-exponential separately   equals   a $Z$-independent
  cohomology term up to terms in  $\Hp$. This  is considered in Section \ref{DVOJNYE}.
The term with degree-six  pre-exponential is considered in Section \ref{TROJNYE}.
As a result of these calculations   $J_6$ \eq{Result5} and $ J_7$ \eq{Result6} are obtained.

  \subsection{Degree-four  pre-exponential}
 \label{DVOJNYE} Consider  the   sum  of expressions with $z$-dependent degree-four
 pre-exponential
from Eqs.~ \eqref{FRest1},  \eqref{FRest2} and \eq{FRest3}, denoting it as $S_4$.
  Partial integration yields
  \bee\label{lostcohomo}&&S_4\approx J_6 +\ff{   \eta^2  }{4 }\int d\Gamma \, \delta(\xi_3)\,%(\p_{\gx_1}-\p_{\gx_2}) \,
 \Big[
   \ff{\gr_2}{(1- \gr_1 -\gr_4)(1-\gr_3)(\gr_1+\gr_4) }
   {t}{}^\ga  z _\ga   ( p_3+p_2)^{\gga}(   {t}-\tilde{t}
    ){}_{\gga}   \\ \nn &&
    +    \ff{ \gr_2 \gr_4}{(1-\gr_1-\gr_4 )(1-\gr_3)^2(\gr_1+\gr_4)^2}
     {t}{}^\gga z_\gga \big(y + \Pz{}\big)^\ga  {t}{} _{\ga}
    \\ \nn&&
   +
     \ff{ \gr_2}{(1-\gr_1-\gr_4 )(1-\gr_3)^2(\gr_1+\gr_4)}
      ( p_1{}+ p_2)^{\gga} \big(y  + (1-\gr_4){t}{}
       \big) _{\gga}   z^\ga  {t}{}_{\ga}
 \\ \nn &&
+        \ff{ \gr_2}{(1- \gr_1 -\gr_4)^2(1-\gr_3)(\gr_1+\gr_4)}  {t}{}^{\ga}z_{\ga}
    \,  ( p_3+p_2)^{\gga} (y+\tilde{t}{})_{\gga}
   \\ \nn && %\label{dB3goB3rest2W1B2sub182}
  %%%%%%%%2222222222222
    +   \ff{ \gr_2}{(1- \gr_1 -\gr_4)^2(1-\gr_3)^2( \gr_1+\gr_4 ) }
\big( -\Pz{}+ \tilde{t}{} \big)^\gga \big( y +   \tilde{t}{}  \big)_{\gga} z^\ga {t}{}_{\ga}
     \Big] \Ee\go CCC\q
%%%%%%%%%======================
\eee
where the cohomology term $J_6$ is given in \eq{Result5}\,.
It is not hard to see that  the
integrand  of the remaining term   is   zero by virtue of the Schouten  identity.

\subsection{Degree-six  pre-exponential}
\label{TROJNYE}

Terms of this type either appear  in \eqref{FRest1}, \eqref{FRest2} via differentiation
   in $\gr_j$ or  in \eqref{FRest3} via  differentiation   in $\gx_j$.
Denoting  a sum of these terms   as $S_6$ we  obtain
%from Eqs.~\eqref{FRest1}, \eqref{FRest2} and \eqref{FRest3}
 \bee\label{SUM3} &&S_6=    +\ff{   \eta^2  }{4 }\int d\Gamma \, \delta(\xi_3)     % \\  \nn&&
 \Big\{
       \Ez (y+ \tilde{t}{} )^{\gga}z_{\gga}\ff{\gr_2}{(1- \gr_1 -\gr_4)(1-\gr_3) }{t}{}^\ga
          (p_1+p_2 ){} _\ga  \Big[
   (\overrightarrow{\p}_{\gr_2}-\overrightarrow{\p}_{\gr_3})E  \Big]\qquad
  \\ \nn&&
   +     \Ez
       \ff{ \gr_2}{(1-\gr_1-\gr_4 )(1-\gr_3)^2}
    \Big[
    %  ( p_1{}+ p_2)^{\gga} z _{\gga}\big(y + \Pz{}\big)^\ga(y+\tilde{t}{})_{\ga}
    ( p_1{}+ p_2)^{\gga} \big(y + (1 -\gr_4 ){t}{}\big)_{\gga}  z_\ga(y+\tilde{t}{})^{\ga}
     \Big]
    [ \overrightarrow{\p}_{\gr_4}-\overrightarrow{\p}_{\gr_1}] E
  \\ \nn&& %2222222222222222222222222
  +  \Ez
     \ff{ \gr_2}{(1-\gr_1-\gr_4 )(1-\gr_3)^2}
       {t}{}^\gga \big(y +  (1-\gr_1-\gr_4 )( p_1{}  +p_2{} ){}\big)_\gga z_\ga(y+\tilde{t}{})^{\ga}
       [\overrightarrow{\p}_{\gr_2}-\overrightarrow{\p}_{\gr_1} ]E
        \\\nn  &&
         %33333333333333333333333333333
+i \gx_1\Big[
    +    \Big\{
   +\ff{ \gr_2\gr_2}{(1- \gr_1 -\gr_4)^3(1-\gr_3)^3( \gr_1+\gr_4 ) }
\big( y+  (1-\gr_1-\gr_4 )( p_1{}  +p_2{} )+  (1 -\gr_4 ){t}{}  \big)^\gga \big( y +   \tilde{t}{}  \big)_{\gga} z_\ga {t}{}^{\ga}
  \\ \nn &&
-\ff{ \gr_3\gr_2}{(1- \gr_1 -\gr_4)^3(1-\gr_3)^3 }
 {\big( y   +    \tilde{t}{} \big)^\gga z_{\gga}   {t}{}^{\ga} y_{\ga}}
  \\ \nn &&
  +    \ff{ \gr_2}{(1- \gr_1 -\gr_4)^2(1-\gr_3)^3 } \big( y
   +    \tilde{t}{} \big)^\gga z_{\gga}       { (p_1 +p_2 )^{\ga}{t}{}_{\ga}}
 \Big\} \Ee\Big]\times%\\ \nn &&
   \big(y  + \Pz{}   \big)^\ga
 (y+\tilde{t}{})_{\ga}\Big\} \go CCC
 \eee
Recall that the integral measure $\dr \Gamma$\eq{dGamma} contains the factor of  $ \gd(1-\sum_1^3 \gx_i)$.
Hence taking into account the factor of $\gd(\gx_3)$ on the \rhs of \eq{SUM3}  the
 dependence on $\gx_2,\gx_3$   can be eliminated
by the substitution $\gx_2\to 1-\gx_1$, $\gx_3\to 0$. Then we consider
  separately  the terms that contain and do not contain $\xi_1$ in the pre-exponentials.
As shown in Appendix D,  those with $\gx_1$-proportional pre-exponentials  give   $J_7$ \eq{Result6} up to $\Hp$,
while those with
 $\gx_1 $-independent   pre-exponentials  give zero up to $\Hp$.

\section{Conclusion}

In this paper starting from $Z$-dominated expression obtained in \cite{4a3}  the  manifestly
spin-local holomorphic vertex $\Upsilon^{\eta\eta}_{\go CCC}$
in the equation \eqref{zeroform}
  is obtained for the $\go CCC$ ordering.
Besides evaluation the  expression for the vertex,
our analysis illustrates  how $Z$-dominance implies spin-locality.

One of the main technical difficulties towards $Z$-independent expression was uniformization,
that is bringing
the exponential factors to the same form, for all contributions
\eqref{origW1B2}-\eqref{kuku5} with the least amount of new integration parameters
possible. Practically, some part of the uniformization procedure heavily used
the Generalized Triangle identity of Section \ref{SecGTid} playing important role in our analysis.

Let us stress that   spin-locality of the  vertices
 obtained in \cite{4a3} follows from    $Z$-dominance Lemma.
 However the evaluation the explicit spin-local vertex
$\Upsilon^{\eta^2}_{\go CCC}$ achieved in this
paper is  technically  involved.  To derive explicit form of other spin-local vertices
in this and higher orders  a more elegant approach to this problem is
highly desirable.

\section*{Acknowledgments}

We  would like  to thank Mikhail Vasiliev  for fruitful discussions and useful comments on
the manuscript.
We acknowledge  a partial support from  the Russian Basic
Research Foundation Grant   No 20-02-00208.
 The work of OG is partially supported by the  FGU FNC SRISA RAS (theme     0065-2019-0007).

\newcounter{appendix}
\setcounter{appendix}{1}
\renewcommand{\theequation}{\Alph{appendix}.\arabic{equation}}
\addtocounter{section}{1} \setcounter{equation}{0}
 \renewcommand{\thesection}{\Alph{appendix}.}
 \addcontentsline{toc}{section}{\,\,\,\,\,\,\,Appendix A: $B_3^{\eta\eta}$}

 \section*{Appendix A: $B_3^{\eta\eta}$}
\label{AppC}

$B_3^{\eta\eta}$ modulo $\Hp$ terms   from \cite{4a3}  is given by
\be
{B}_3^{\eta\eta}\approx-\frac{\eta^2}{4}  \int d\Gamma  \delta(\xi_3) \delta(\rho_4)\frac{\T\rho_2 (z_\ga y^\ga)^2}{(\rho_1+\rho_2)(\rho_2+\rho_3)}
  \exp\big(\KE \big)CCC
\q\ee
 where $d\Gamma$ is defined  in \eq{dGamma},
\begin{equation}
\KE=i\T z_\ga\left(y^\ga+\PP_0^\ga\right)
+\frac{i(1-\xi_1) \rho_2}{\rho_1+\rho_2}y^\ga (p_{1\ga}+p_{2\ga})
+\frac{i\xi_1 \rho_2}{\rho_2+\rho_3}y^\ga (p_{2\ga}+p_{3\ga})-iy^\ga p_{2\ga}\q
\end{equation}
 \begin{equation}
\PP_0=(1-\rho_1)(p_1+p_2)-(1-\rho_3)(p_2+p_3).
\end{equation}
Performing partial integration with respect to $\T$ twice we obtain
\be\label{B31}
{B}_3^{\eta\eta}\approx\frac{\eta^2}{4}  \int d\Gamma\frac{\delta(\xi_3)\delta(\rho_4)\rho_2}{(1-\rho_3)(1-\rho_1)}
  \Big[\delta(\T)+iz_\ga \PP_0^\ga+iz_\ga \PP_0^\ga
\Big(1+i\T z_\ga \PP_0^\ga\Big)\Big]   \exp\big( \KE\big)CCC
\,.
\ee Noticing that
\begin{equation}
\frac{\p}{\p \rho_1 } \KE
=-i\T z_\ga (p_1 {}^\ga+p_2 {}^\ga)-i\frac{(1-\xi_1)\rho_2}{(\rho_1+\rho_2)^2}y^\ga(p_{1\ga}+p_{2\ga}),
\end{equation}
\begin{equation}
\frac{\p}{\p \rho_3} \KE=\\
=i\T z_\ga (p_2 {}^\ga + p_3 {}^\ga)-i\frac{\xi_1 \rho_2}{(\rho_2+\rho_3)^2}
y^\ga (p_{2\ga}+p_{3\ga})
\end{equation}
and
performing partial integration    with respect to $\rho_1$ and $\rho_3$ we obtain
\begin{multline}
{B}_3^{\eta\eta}\approx\frac{i\eta^2}{4}\int d\Gamma
\frac{\delta(\xi_3)\delta(\rho_4)}{(1-\rho_3)(1-\rho_1)}
\Bigg[ {-i\rho_2
\delta(\T)}
 + \,  z_\ga \PP_0^\ga \big({(1-\rho_3)(1-\rho_1)}\left(\delta(\rho_1)+\delta(\rho_3)\right)
-1\big)    \\
-  { i\,\rho_2 z_\ga
\PP_0^\ga} \left(  \xi_2 \frac{y^\ga(p_{1\ga}+p_{2\ga})}{(\rho_1+\rho_2)}
+ \xi_1 \frac{y^\ga(p_{2\ga}+p_{3\ga})}{(\rho_2+\rho_3)}\right)\Bigg]\exp\big(\KE \big) CCC.
\end{multline}
Observing that
\begin{equation}
\frac{\p \KE}{\p \xi_1}=\frac{i\rho_2}{\rho_2+\rho_3} y^\ga (p_{2\ga}+p_{3\ga})-\frac{i\rho_2}{\rho_1+\rho_2} y^\ga (p_{1\ga}+p_{2\ga})
\end{equation}
and using the Schouten identity
\begin{equation}
z_\ga (p_2 {}^\ga+p_3 {}^\ga) y^\beta (p_{1\beta}+p_{2\beta})=z_\ga y^\ga (p_2 {}^\beta +p_3 {}^\beta)(p_{1\beta}+p_{2\beta})+z_\ga (p_1 {}^\ga+p_2 {}^\ga) y^\beta(p_{2\beta}+p_{3\beta})
\end{equation}
 after partial integration with respect to $\xi_1$ we obtain
\begin{multline}\label{B3modH=1406}
{B}_3^{\eta\eta}\approx\frac{i\eta^2}{4}\int d\Gamma
\frac{\delta(\xi_3)\delta(\rho_4)}{(1-\rho_3)(1-\rho_1)}
\Bigg[ {-i\rho_2
\delta(\T)}+   {z_\ga(p_1 {}^\ga+p_2 {}^\ga)(1-\rho_1)} \Big(\delta(\xi_2)-\delta(\xi_1)\Big)
\\
+  z_\ga \PP_0^\ga \Big[(1-\rho_1 )(1-\rho_3)\Big(\delta(\rho_1)
+\delta(\rho_3)\Big)-\delta(\xi_2) \xi_1\Big]
+i\rho_2 z_\ga y^\ga (p_{1\beta}+p_{2\beta})(p_2 {}^\beta+p_3 {}^\beta)
\Bigg]\exp\big(\KE \big) CCC.
\end{multline}
The $\delta(\T)$-proportional term gives rise to $J_1$ \eq{go B3modHcoh} and $J_2$ \eq{dx B3modHcoh}.

\addtocounter{appendix}{1}
\renewcommand{\theequation}{\Alph{appendix}.\arabic{equation}}
\addtocounter{section}{1} \setcounter{equation}{0}
 \addcontentsline{toc}{section}{\,\,\,\,\,\,\,Appendix B: Details of uniformization}

\section*{Appendix B: Uniformization  Detail }

\label{Auniform}
Here  some  details of the transformation of
integrands  \eqref{RRwB3modH+}--\eqref{FFFFFFFFk=}\, to the  form \eq{comexp} are   presented.

Uniformization can be easily achieved  for    Eqs.~\eq{RRwB3modH+} and \eq{RRdxB3modH+} modulo $\gd(\gr_1)$-proportional terms.
 Indeed,   eliminating $\gd(\gr_1)$-proportional term from the \rhs  of  \eq{RRwB3modH+}, adding an integration parameter
 $ \gr_4  $ and a factor of $\gd(\gr_4 )$,
one  obtains \eqref{F1}.
Analogously, eliminating $\gd(\gr_1)$-proportional term from the \rhs \eq{RRdxB3modH+},
adding an integration parameter
 $ \gr_4  $,
swapping $ \gr_1\leftrightarrow \gr_4$ and  then adding
    a factor of $\gd(\gr_1 )$
one obtains \eqref{F2}.

To transform integrands of Eqs.~\eq{W2C3gr1} and \eq{FFFFFFFFk=}, as well as
%some parts of
$\gd(\gr_1)$-proportional terms    of the integrands of  Eqs.~\eq{RRwB3modH+} and \eq{RRdxB3modH+},
to the
  form  \eq{comexp}
      GT identity  \eq{GTH+F} is  used in Sections {  B.1} and {  B.2}.

  \subsection{ $d_x B_2 {} + W_2 * C   $}
  \label{GTdxB2+}

Noticing that  the  exponential of \eqref{W2C3gr1}  coincides with $\Ee$  at $\xi_2=0$, while the exponential of \eqref{FFFFFFFFk=} coincides with    $\Ee$ \eqref{Ee}
 at  $\xi_1=0$,
 one can easily make sure, that
  only
 the $\gd(\gx_2)$-proportional term   of
\eq{W2C3gr1} and the $\gd(\gr_1)$-proportional term   of  \eq{FFFFFFFFk=} have  the desired
   form \eq{comexp}.

Using that $\Ee$ \eqref{Ee} does not depend on $\gx_3$, swapping $\gx_3 \leftrightarrow \gx_1$ in
the remaining part of \eq{FFFFFFFFk=}, then swapping $\gx_3 \leftrightarrow \gx_2$
in the remaining part of \eq{W2C3gr1}, one then can apply GT identity \eqref{GTH+==0} to the sum of the
two obtained
 terms .
 As a result,       Eqs.~\eqref{W2C3gr1}, \eqref{FFFFFFFFk=}
yield
 \bee\label{D4}&&
\dr_x B_2^{\eta\, loc}+W_{2\, \go CC}^{\eta\eta}\ast C\approx\frac{\eta^2}{4}\int d\Gamma\, \delta(\rho_3)\delta(\xi_3)
\Big[-i\frac{(z_\ga t^\ga)}{\rho_1+\rho_4}
\delta(\xi_2)-i(\underline{z_\ga y^\ga}) \delta(\rho_1)\Big] \mathcal{E} \go CCC\qquad\\\nn&&+
\frac{\eta^2}{4}
\int d\Gamma\, \delta(\rho_3)\Big[i\delta(\rho_4)-t^\gga(p_{1\gga}+p_{2\gga})\Big]
\Big\{\delta(\T) \widetilde{t}^\ga y_\ga
 + \delta(\xi_3)
 (z_\ga \widetilde{t}^\ga+\underline{z_\ga y^\ga})\Big\}\mathcal{E} \go CCC\q
\eee
where the   terms in the second row of formula \eq{D4} result from applying $GT$ -identity.
Rewriting the underlined part as the result of differentiation with respect to $\T$ and
performing  partial integration  one obtains Eq.~\eqref{F3} plus the cohomology term  $J_8$
 \eqref{Result1}.

\subsection{  $(\dr_x B^{\eta\eta}_3{}+\go* B^{\eta\eta}_3)|_{\gd(\gr_1)}+W^\eta_{1\, \go C}*B^{\eta\, loc}_2$}

%\subsubsection{GT-transformation of $(d_x B_3{}+\go* B_3)|_{\gd(\gr_1)}$ }
    \label{GTB3des}
%Though $\gd(\gr_1)-$proportional terms on the \rhss of \eq{RRdxB3modH+} and \eq{RRwB3modH+} cannot be
%straightforwardly uniformized, this .

Uniformization of the sum of $\gd(\gr_1)-$proportional terms on the \rhss of \eq{RRdxB3modH+} and \eq{RRwB3modH+}
is done with the help of $GT$ identity  \eqref{GTH+==0} as follows.
Denoting \be
\widetilde{P}=    y+    p_1+p_2+{t} -  \gr_2  (p_3+p_2)
\ee
one can see that partial integration in $\T$ yields
\begin{multline}\label{D6}
\dr_x {B}^{\eta\eta}_3 \bigg|_{\delta(\rho_1)}\approx-\frac{i\eta^2}{4}\int d\Gamma\,
\delta(\rho_4)\delta(\rho_1)\delta(\xi_1) \Big[i\delta(\T)-z_\ga y^\ga\Big]
\exp\Big\{i\T z_\ga \widetilde{P}^\ga-i\xi_2 \widetilde{P}^\ga y_\ga \\
+i(1-\rho_2)(p_2 {}^\ga +p_3 {}^\ga)y_\ga+ip_{3\ga} y^\ga+it^\beta p_{1\beta} \Big\}\go CCC,
\end{multline}
\begin{multline}\label{D7}
\go\ast {B}^{\eta\eta}_3\bigg|_{\delta(\rho_1)}\approx \frac{i\eta^2}{4}\int d\Gamma\, \delta(\rho_4)\delta(\rho_1)\delta(\xi_3)\Big[i\delta(\T)-z_\ga(y^\ga+t^\ga)\Big]\exp\Big\{i\T z_\ga \widetilde{P}^\ga-i\xi_2 \widetilde{P}^\ga y_\ga \\
+i\xi_1 \widetilde{P}^\ga t_\ga+i(1-\rho_2)(p_2 {}^\ga +p_3 {}^\ga)y_\ga+ip_{3\ga} y^\ga+it^\beta p_{1\beta} \Big\}\go CCC\,.
\end{multline}

The sum of \eqref{D6} and \eqref{D7}   gives
\begin{multline}
\Big(\dr_x {B}^{\eta\eta}_3+\omega \ast B_3^{\eta\eta}\Big)\bigg|_{\gd(\rho_1)} \approx
  \frac{i\eta^2}{4}\int d\Gamma\,\gd(\rho_4)\gd(\rho_1)\Big[z_\gga(-t^\gga-y^\ga)\gd(\xi_3)+z_\gga y^\gga\gd(\xi_1)+z_\gga t^\gga\gd(\xi_2)\Big]\times\\
\times \exp\Big\{i\T z_\ga \widetilde{P}^\ga-i\xi_2 \widetilde{P}^\ga y_\ga
+i\xi_1 \widetilde{P}^\ga t_\ga+i(1-\rho_2)(p_2 {}^\ga +p_3 {}^\ga)y_\ga+ip_{3\ga} y^\ga+it^\gb p_{1\gb} \Big\} \go CCC \\
-\frac{i\eta^2}{4}\int d\Gamma\, \gd(\rho_4)\gd(\rho_1)
(z_\gga t^\gga)\gd(\xi_2)\exp\Big\{i\T z_\ga \widetilde{P}^\ga-i\xi_2 \widetilde{P}^\ga y_\ga
+i\xi_1 \widetilde{P}^\ga t_\ga+i(1-\rho_2)(p_2 {}^\ga +p_3 {}^\ga)y_\ga \\
+ip_{3\ga} y^\ga+it^\gb p_{1\gb}\Big\}\go CCC \,+ J_9+J_{10}
   \label{ERRGT}\end{multline}
   with $J_9$ \eq{goB3modH1406gr1C} and $J_{10}$ \eq{dxB3modH1406gr1C}.
By virtue of GT identity \eqref{GTH+==0} the first term weakly equals   $J_{11}$ \eq{ERRGTC}.
  Finally,    Eq.~\eq{ERRGT}  yields
% JJJ
\begin{multline} \label{dB3+wB3}
\Big(\dr_x {B}^{\eta\eta}_3+\omega \ast B_3^{\eta\eta}\Big)\bigg|_{\gd(\rho_1)} \approx
-\frac{i\eta^2}{4}\int d\Gamma\, \gd(\rho_4)\gd(\rho_1) (z_\gga t^\gga)\gd(\xi_2)\exp\Big\{i\T z_\ga \widetilde{P}^\ga-i\xi_2 \widetilde{P}^\ga y_\ga
 \\
+i\xi_1 \widetilde{P}^\ga t_\ga+i(1-\rho_2)(p_2 {}^\ga +p_3 {}^\ga)y_\ga+ip_{3\ga} y^\ga+it^\gb p_{1\gb}\Big\}
\go CCC\, + J_9+J_{10}+J_{11}.
\end{multline}
Consider $W_{1 \go C}^\eta \ast B_2^{\eta\, loc}$ \eq{origW1B2}.
This is convenient  to change  integration variables,
moving from the integration over simplex to  integration over square. As a result
  \begin{multline}\label{W1B2mod1}
W_{1 \go C}^\eta \ast B_2^{\eta\, loc}\approx \frac{\eta^2}{4}\int_0^1 d\T\, \T \int d^2 \tau_+\, \gd(1-\tau_1-\tau_2)\int_0^1 d\sigma_1 \int_0^1 d\sigma_2\, (z_\ga t^\ga)\times \\
\Big[z_\ga y^\ga+\sigma_1 z_\ga t^\ga\Big]\exp\Big\{i\T z_\ga y^\ga+i(1-\sigma_2)\sigma_1 t_\ga p_1 {}^\ga+i\sigma_1\sigma_2 t^\ga p_{3\ga}+i(1-\sigma_1)t^\ga p_{1\ga} \\
+i\T z_\ga \Big((\tau_1+\tau_2 \sigma_1)t^\ga+\tau_1 p_1 {}^\ga-(\tau_2-\tau_1(1-\sigma_2))p_2 {}^\ga-(\tau_2+\sigma_2\tau_1)p_3 {}^\ga\Big)+i\sigma_1 y^\ga t_\ga \\
-i(1-\sigma_2)y^\ga p_{2\ga}+i\sigma_2 y^\ga p_{3\ga}+i\sigma_2 y^\ga p_{3\ga} \Big\}\go CCC.
\end{multline}  Partial integration with respect to $\T$
yields\begin{multline}
W_{1 \go C}^\eta \ast B_2^{\eta\, loc}\approx-\frac{\eta^2}{4}\int_0^1 d\T \int d^2 \tau_+\,
\gd(1-\tau_1-\tau_2)\int_0^1 d\sigma_1 \int_0^1 d\sigma_2\, (z_\ga t^\ga)\times \\
\Big[\T z_\ga \Big(\tau_1(p_1 {}^\ga+p_2 {}^\ga)-(\tau_2+\sigma_2 \tau_1)(p_2 {}^\ga +p_3 {}^\ga)\Big)
-i\T \tau_1 (1-\sigma_1) z_\ga t^\ga\Big]\,\exp(\KEE)\,\,\go CCC\q
\end{multline}where
\begin{multline}\label{tildeEe}
\KEE=i\T z_\ga y^\ga+it^\gb p_{1\gb}+i\sigma_1\Big(y^\ga t_\ga+(p_1 {}^\ga+p_2 {}^\ga)t_\ga
-\sigma_2(p_2 {}^\ga+p_3 {}^\ga)t_\ga\Big)-i\big(\sigma_2 p_3 {}^\ga-(1-\sigma_2)p_2 {}^\ga\big)y_\ga \\
+i\T z_\ga \Big(\tau_1(p_1 {}^\ga +p_2 {}^\ga)-(\tau_2+\sigma_2\tau_1)(p_2 {}^\ga+p_3 {}^\ga)
+(\sigma_1+\tau_1(1-\sigma_1))t^\ga\Big).
\end{multline}
By virtue of evident formulas
\bee\nn&&
\tau_1 \left(\frac{\p}{\p \tau_1}-\frac{\p}{\p \tau_2}\right)
\KEE=i\T z_\ga \Big(\tau_1(p_1  +p_2 {} )+\big[(\tau_1+\tau_2)-(\tau_2+\sigma_2\tau_1)\big]
(p_2 {} +p_3 {} )+\tau_1(1-\sigma_1)t  \Big){}^\ga \q
\\ \nn&&\frac{\p}{\p \sigma_1}\KEE=
i\T (1-\tau_1)z_\ga t^\ga+i\Big(y^\ga+p_1 {}^\ga+p_2 {}^\ga-\sigma_2 (p_2 {}^\ga+p_3 {}^\ga)\Big)t_\ga,
\eee
 Eq.~\eqref{W1B2mod1} acquires  the form
\begin{multline}
W_{1 \go C}^\eta \ast B_2^{\eta\, loc}\approx\frac{\eta^2}{4}\int_0^1 d\T\int d^2
\tau_+\gd(1-\tau_1-\tau_2)\int_0^1 d\sigma_1 \int_0^1 d\sigma_2\bigg[iz_\ga t^\ga \tau_1
\left(\frac{\p}{\p \tau_1}-\frac{\p}{\p \tau_2}\right) \\
-\frac{z_\ga (p_2 {}^\ga+ p_3 {}^\ga)}{1-\tau_1}\left(i\frac{\p}{\p \sigma_1}
+\Big(y^\ga+p_1 {}^\ga+p_2 {}^\ga-\sigma_2(p_2 {}^\ga+p_3 {}^\ga)\Big)t_\ga\right)+iz_\ga t^\ga\bigg]
\exp(\KEE )\go CCC.
\end{multline}
After partial integrations   in $\tau_1$,$\tau_2$ and $\sigma_1$ one obtains
\bee&&\label{underlC10}
W_{1 \go C}^\eta \ast B_2^{\eta\, loc}\approx\frac{\eta^2}{4}\int_0^1
d\T\int d^2 \tau_+\gd(1-\tau_1-\tau_2)\int_0^1 d\sigma_1 \int_0^1 d\sigma_2
\bigg[\underline{iz_\ga t^\ga \gd(\tau_2)} \\
&&\nn+\frac{z_\ga (p_2 {}^\ga+ p_3 {}^\ga)}{1-\tau_1}\left(i\big(\gd(\sigma_1)-\gd(1-\sigma_1)\big)
-\Big(y^\ga+p_1 {}^\ga+p_2 {}^\ga-\sigma_2(p_2 {}^\ga+p_3 {}^\ga)\Big)t_\ga\right)\bigg]
\exp (\KEE )\go CCC\,.
\eee
After a simple change of integration variables the underlined term on the \rhs of Eq.~\eq{underlC10}
  cancels  the \rhs of  Eq.~\eqref{dB3+wB3}. Performing integration    with respect to $\gt_2$
   in the remaining part of \eq{underlC10},
after  the following change of the integration variables
\bee\nn&&
\int_0^1 d\sigma_1\int_0^1 d\gt_1 \int_0^1 d\sigma_2\, f(\sigma_1,1-\sigma_1,\gt_1,\sigma_2)\\\nn&&
=\int d^4 \rho_+\, \delta\left(1-\sum_{j=1}^4 \rho_j\right)\frac{1}{(\gr_2+\gr_3)(1-\gr_2-\gr_3)}
f\left(\frac{\gr_1}{1-\gr_2-\gr_3},\frac{\rho_4}{1-\gr_2-\gr_3},\gr_2+\gr_3,\frac{\gr_2}{\gr_2+\gr_3}\right)
\,, \eee
  $\exp(\KEE)$ \eq{tildeEe} acquires the form $\Ee$ \eq{Ee}. As a result,  the sum of    Eq.~\eq{underlC10} and Eq.~\eqref{dB3+wB3}   by virtue  Eq.~\eqref{EEgx14=}
yields Eq.~\eqref{F4}.

\addtocounter{appendix}{1}
\renewcommand{\theequation}{\Alph{appendix}.\arabic{equation}}
\addtocounter{section}{1} \setcounter{equation}{0}
\setcounter{subsection}{0}
 \addcontentsline{toc}{section}{\,\,\,\,\,\,\,Appendix C:  Eliminating    $\gd(\gr_j)$ and $\gd(\gx_j)$}
 \renewcommand{\thesubsection}{\Alph{appendix}.\arabic{subsection}}

\section*{Appendix C:   Eliminating    $\gd(\gr_j)$ and $\gd(\gx_j)$}
\label{AppD}
To eliminate $\gd(\gr_j)$ and $\gd(\gx_j)$ from of the \rhss of Eqs.~\eqref{F1}, \eqref{F2}
this is convenient to group      similar pre-exponential terms as in  Sections \ref{Eli1} -\ref{Eli4}.
   \subsection{Terms proportional to $( p_1{}+ p_2)^{\ga}   ( p_3{}+p_2)_{\ga}$}
\label{Eli1}

 Consider
    $F_{1,1}+F_{2,1}$ of \eqref{F1} and \eqref{F2}, respectively.
   Partial integration  with respect to $\rho_1$ and $\rho_4$ yields
   % delta-functions $\delta(\rho_1)$ and $\delta(\rho_4)$  yields
\begin{multline}\label{F11+F21b}
F_{1,1}+F_{2,1}\approx - \frac{\eta^2}{4}\int d\Gamma\frac{\delta(\xi_3)\rho_2}{(1-\rho_1-\rho_4)(1-\rho_3)}(p_1 {}^\ga+p_2 {}^\ga)(p_{2\ga}+p_{3\ga}) \times \\
\times (z_\gga \PP^\gga)\left(\frac{\p}{\p \rho_4}
-\frac{\p}{\p \rho_1}\right)\Ee \go CCC.
\end{multline}
By direct calculation, Eq.~\eqref{F11+F21b} gives
\begin{multline}
F_{1,1}+F_{2,1}\approx -\frac{\eta^2}{4}\int d\Gamma\frac{\delta(\xi_3)\rho_2}{(1-\rho_1-\rho_4)(1-\rho_3)}(p_1 {}^\ga+p_2 {}^\ga)(p_{2\ga}+p_{3\ga})\times\\
\Bigg[\Ez \left(\frac{\p}{\p \rho_4}-
\frac{\p}{\p \rho_1}\right) (z_\gga \PP^\gga) E+(z_\gga \PP^\gga)
\T (z_\ga t^\ga)\mathcal{E} \Bigg]\go CCC\,.
\end{multline}
By virtue of the Schouten identity
\begin{equation}
 z_\ga t^\ga   (p_1+p_2 )^\gga  ( p_3{}  +p_2{} )_\gga=
 t^\ga(p_1+p_2 ){} _\ga    z^\gga  ( p_3{}  +p_2{} )_\gga+ t^\ga ( p_3{}  +p_2{} )_\ga     (p_1+p_2 )^\gga  z   _\gga
\end{equation}
and its  consequence
\begin{multline}\label{SchCons}
 z_\ga t^\ga   (p_1+p_2 )^\gga  ( p_3{}  +p_2{} )_\gga \Ee%=\\
=t^\ga(p_{1 }+p_{2 })_\ga\left[i\left(\frac{\overleftarrow{\p}}{\p \rho_2}
-\frac{\overleftarrow{\p}}{\p \rho_3}\right)\Ez E+i\Ez
\left(\frac{\p}{\p \rho_2}-\frac{\p}{\p \rho_3}\right)E\right] \\
+t^\ga (p_{2 }+p_{3 })_\ga\left[i\left(\frac{\overleftarrow{\p} }
{\p \rho_2}-\frac{\overleftarrow{\p}}{\p \rho_1}\right)\Ez E
+i\Ez \left(\frac{\p}{\p \rho_2}-\frac{\p}{\p \rho_1}\right) E\right]\,
\end{multline}
Eq.~\eqref{F11+F21b} yields
\begin{multline}\label{F11+F21}
F_{1,1}+F_{2,1}\approx + \frac{\eta^2}{4}\int d\Gamma\, \delta(\xi_3) \Bigg\{\ff{(z_{\gga}\PP ^{\gga})\gr_2}{(1- \gr_1 -\gr_4)(1-\gr_3) }\times\\
\times\Bigg(( p_1{}+ p_2)^{\ga}   ( p_3{}+p_2)_{\ga}
    \Ez   \Bigg[\frac{\p}{\p \rho_1}-\frac{\p}{\p \rho_4}\Bigg]  E
  +t^\ga (p_1+p_2 ){} _\ga   \Bigg[  \underline{ \gd(\gr_3)} \Ee
  - \Ez\Bigg(\frac{\p}{\p \rho_2}-\frac{\p}{\p \rho_3}\Bigg)E  \Bigg] \\
+t^\ga ( p_3{}  +p_2{} )_\ga   \Bigg[ \underline{ \gd(\gr_1)}\Ee
  - \Ez \Bigg(\frac{\p}{\p \rho_2}-\frac{\p}{\p \rho_1}\Bigg)E  \Bigg]
\Bigg)+\ff{\gr_2}{(1- \gr_1 -\gr_4)(1-\gr_3) }
\Big(  t^\ga  z _\ga   ( p_3+p_2)^{\gga}(p_1+p_2 ){}_{\gga}  \Ee
    \Big)\\
+(z_{\gga}\PP ^{\gga})\Bigg(- \ff{1-\gr_3-\gr_2}{(1- \gr_1 -\gr_4)(1-\gr_3)^2 } t^\ga  (p_1+p_2 ){} _\ga
 \Ee- \ff{1- \gr_1 -\gr_4-\gr_2}{(1- \gr_1 -\gr_4)^2(1-\gr_3) }t^\ga ( p_3{}  +p_2{} )_\ga
 \Ee \Bigg)\Bigg\}\go CCC\,.
\end{multline}
One can see that $\delta(\rho_1) $- and
$\delta(\rho_3) $-proportional terms on the \rhs of \eq{F11+F21} (the underlined ones)
cancel  terms  $F_{2,4}$   \eqref{F2} and $F_{3,3}$   \eqref{F3}, respectively.

\subsection{Term proportional to $t^\ga(p_{1\ga}+p_{2\ga})$}
Consider   term  $F_{3,5}$ of $F_{3 }$ \eqref{F3}. By virtue of the following identity
\begin{equation}
\frac{\rho_2}{(\rho_2+\rho_3)(1-\rho_3)}\left(\delta(\rho_3)-\delta(\rho_2)\right)=1
\end{equation}
\begin{multline}
F_{3,5}  \approx -\frac{\eta^2}{4}\int d\Gamma\, \frac{\delta(\xi_3)\rho_2}{(\rho_2+\rho_3)(1-\rho_3)}\Big(\delta(\rho_3)-\delta(\rho_2)\Big)
\\  \Big[  ( p_2{}_\ga+ p_1{}_\ga) t^{\ga}
    (z_{\gga}t^\gga)\Big(  (1-\gr_4)  -\ff{\gr_1}{(\gr_1+\gr_4)} \Big)\Ee  \Big]\go CCC.
\end{multline}
Partial integrations along with the Schouten identity
\begin{equation}
t^\ga(p_{1\ga}+p_{2\ga} )   ( p_3{}^\gga  +p_2{}^\gga )  z_\gga
= - \underline{t^\ga z _\ga}   (p_1{}^\gga+p_2{}^\gga )  ( p_{2\gga}+p_{3\gga})
+ t^\ga ( p_{3\ga}  +p_{2\ga} )  \underline{ (p_1+p_2 )^\gga  z   _\gga}
\end{equation}
and realization of the underlined  terms as derivative of $\Ez$
 along with  further partial integration
yields
\begin{multline}\label{F35=}
F_{3,5}\approx -\frac{\eta^2}{4}\int d\Gamma\, \delta(\xi_3)\Bigg[\ff{ \gr_4}{ (1-\gr_3)^2} \Big(  ( p_2{}_\ga+ p_1{}_\ga) t^{\ga} z_{\gga}t^\gga    \Big)\Ee \\
+\ff{\gr_2\gr_4}{(\gr_1+\gr_4)(1-\gr_3)} \Big(   ( p_2{}_\ga+ p_1{}_\ga) t^{\ga}z_{\gga}t^\gga   \Big)\Ez\left[\frac{\p}{\p \rho_2}-\frac{\p}{\p \rho_3}\right]E+ \ff{\gr_2\gr_4}{(\gr_1+\gr_4)(1-\gr_3)} (z_{\ga}t^\ga) \times\\
\times\Bigg( -  (p_1+p_2 )^\gga  ( p_3{}  +p_2{} )_\gga \Ez  \Bigg[\frac{\p}{\p \rho_1}-\frac{\p}{\p \rho_4}\Bigg]E
- \underline{\gd(\gr_1)} (p_1+p_2 )^\gga  ( p_3{}  +p_2{} )_\gga\Ee\\
 -  t^\ga( ( p_3{}  +p_2{} )_\ga  )\Ez  \Bigg[\frac{\p}{\p \rho_1}-\frac{\p}{\p \rho_2}\Bigg]E
- \underline{\gd(\gr_1)}t^\ga( ( p_3{}  +p_2{} )_\ga  )\Ee\Bigg)\\
+(z_{\ga}t^\ga)\Bigg(
\ff{ \gr_2}{ (1-\gr_3) (\gr_1+\gr_4)} (p_1+p_2 )^\gga  ( p_3{}  +p_2{} )_\gga\Ee
  +  \ff{\gr_4 }{(\gr_1+\gr_4)^2} t^\ga( ( p_3{}  +p_2{} )_\ga  )\Ee
\Bigg) \Bigg]\go CCC.
\end{multline}
One can see that the sum of the underlined $\delta(\rho_1)$-proportional terms   cancel   $F_{2,2}+F_{2,3}$  of \eqref{F2}.

\subsection{Sum of $( p_1{}+ p_2)^{\ga}   ( p_3{}+p_2)_{\ga}$-proportional and
 $t^\ga(p_{1\ga}+p_{2\ga})$--proportional terms}

Summing up  $F_{1,1}+F_{2,1} $ \eqref{F11+F21},
      $F_{3,3}$  \eqref{F3},
$F_{3,5}$ \eqref{F35=}  and  $F_{2,2}+F_{2,3}+F_{2,4}$    \eqref{F2}, then performing partial integrations
and  using the following simple identities
\begin{equation}
(1-\gr_4)  -\ff{\gr_1}{(\gr_1+\gr_4)}=
 \ff{\gr_4( \gr_2+\gr_3)}{(\gr_1+\gr_4)},
\end{equation}
\begin{equation}
- \ff{\gr_4 }{(\gr_1+\gr_4)^2} + \ff{\gr_4}{(   \gr_1+\gr_4)}
\ff{ \gr_3}{(1- \gr_1 -\gr_4) (1-\gr_3) }
=  \ff{-\gr_2\gr_4 }{(\gr_1+\gr_4)^2(1- \gr_1 -\gr_4) (1-\gr_3)}\q
\end{equation}
one obtains by virtue of  Eqs.~\eq{tildet}-\eq{PP=}  %(in notations of Section \ref{uniform})
\be \label{FRest1=}
 F_{1,1}+F_{2,1}+F_{2,4}+F_{3,3}+F_{3,5}+F_{2,2}+F_{2,3}=G_1 \ee
 with $G_1$   \eq{FRest1}.

 \subsection{Terms proportional to $\delta(\xi_1)-\delta(\xi_2)$}
\label{Eli3}

Consider a sum of
$F_{1,4}$   \eqref{F1} and $F_{2,8}$ \eqref{F2}.
Performing partial integrations   with respect to $\rho_1$ and $\rho_4$, then applying the Schouten identity
 one obtains
\begin{multline}\label{F14+F28}
 F_{1,4}+F_{2,8}\approx -\frac{\eta^2}{4}\int d\Gamma \, \delta(\xi_3)
 \Bigg[\frac{\p}{\p \rho_1}-\frac{\p}{\p \rho_4}\Bigg]\frac{i z_\ga (p_1 {}^\ga+p_2 {}^\ga)}{1-\rho_3}
 \Big(\delta(\xi_2)-\delta(\xi_1)\Big)\Ee \,\go CCC=\\
=-\frac{\eta^2}{4}\int d\Gamma \,
\delta(\xi_3)\Big(\delta(\xi_2)-\delta(\xi_1)\Big)\Bigg\{\frac{i\, z_\gga t^\gga}{(1-\rho_3)}
\Bigg(\Ez\Bigg[\frac{\p}{\p \rho_1}-\frac{\p}{\p \rho_2}\Bigg]E+\Big(\underline{\delta(\rho_1)}
-\underline{\underline{\delta(\rho_2)}}\Big)\Ee\Bigg) \\
+\frac{i\, z_\ga (p_1 {}^\ga+p_2 {}^\ga)}{(1-\rho_3)}\Ez
\Bigg[\frac{\p}{\p \rho_1}-\frac{\p}{\p \rho_4}\Bigg]E\Bigg\}\go CCC.
\end{multline}

The underlined $ \gd(\gr_1)$-proportional   term compensates $F_{2,9}$  of \eqref{F2}.
The double underlined $\gd(\gr_2)$-proportional  term   vanishes due to the factor of $(\gd(\gx_2)-\gd(\gx_1))$
which after partial integrations  in $\xi_1$ and $\xi_2$ produces an expression
proportional to $\rho_2$.

Summing up  $F_{1,4}+F_{2,8}$ \eq{F14+F28}  and $F_{2,9}$   \eqref{F2},
performing  partial integrations with respect to  $\gx$ and $\T$ along with the  Schouten  identity
one obtains \be \label{FRest2G}
 F_{1,4}+F_{2,8}+F_{2,9}\approx G_2
\ee
with $G_2$  \eq{FRest2}.

 \subsection{Terms proportional to $\xi_1 \delta(\xi_2)$ }
\label{Eli4}

Consider a sum of      $F_{1,3}$ \eqref{F1},  $F_{2,6}$ \eqref{F2} and  $F_{4,1}$ \eqref{F4}.
\bee
F_{1,3}+F_{2,6}+F_{4,1}\approx \frac{i\eta^2}{4}
\int d\Gamma\,\ff{ \delta(\xi_3)\delta(\xi_2)[\delta(\rho_1)-\delta(\rho_4)]}{(\rho_2+\rho_3)} z_\ga
\bigg\{\frac{\PP^\ga
}{(1-\rho_3)}
%\\\nn
- \frac{\xi_1 \,(p_2 {}^\ga+p_3 {}^\ga)}{
(\rho_1+\rho_4)}
\bigg\}\Ee\, \go CCC.\quad
\eee
Partial integration  yields
\begin{multline}\label{FRest2_5-}
F_{1,3}+F_{2,6}+F_{4,1}\approx\frac{i\eta^2}{4}\int d\Gamma\,
\gd(\xi_3)\gd(\xi_2)\xi_1 \Bigg\{z_\ga t^\ga\Bigg[\frac{1}{\rho_1+\rho_4}
\bigg(\Ez\bigg[\frac{\p}{\p \rho_2}-\frac{\p}{\p \rho_3}\bigg]E+\Big[\underline{\gd(\rho_2)}
-\gd(\rho_3)\Big]\Ee\bigg)
 \\
+\frac{1}{1-\rho_3}\bigg(\Ez \bigg[\frac{\p}{\p \rho_1}-\frac{\p}{\p \rho_2}\bigg]E
+\Big[\gd(\rho_1)-\underline{\gd(\rho_2)}\Big]\Ee\bigg)\Bigg] \\
+\bigg[\frac{z_\ga(p_2 {}^\ga+p_3 {}^\ga)}{\rho_1+\rho_4}
+\frac{z_\ga(p_1 {}^\ga+p_2 {}^\ga)}{1-\rho_3}\bigg]\Ez
\bigg[\frac{\p}{\p \rho_1}-\frac{\p}{\p \rho_4}\bigg]E\Bigg\}\go CCC
\,.\end{multline}
One can see that the underlined  $\gd({\gr_2})$-proportional    terms vanish
due to the factor of $\gd(1-\sum\gr_i)$ \eq{dGamma}, while $\gd({\gr_1})$-proportional term compensates
 $F_{2,7}$
\eqref{F2} and  $\gd({\gr_3})$-proportional term
compensates $F_{3,2}$    \eqref{F3}.

Summing up $F_{2,7}$  \eqref{F2},    $F_{3,2}$ \eqref{F2},  $F_{4,2}$  and
$F_{1,3}+F_{2,6}+F_{4,1}$ \eqref{F4},
and then
performing partial integration in $\T$ one obtains   by virtue of the Schouten
identity
\begin{multline}\label{G3=}
F_{1,3}+F_{2,6}+F_{4,1}+F_{2,7}+F_{3,2}+F_{4,2}\approx G_3:=\frac{\eta^2}{4}\int d\Gamma\, \delta(\xi_3)\delta(\xi_2)\times\\
\times\Bigg\{\frac{\rho_2\, (z_\ga t^\ga)(p_2 {}^\gga+p_3 {}^\gga)(y_\gga
+\tilde{t}_\gga)}{(1-\rho_1-\rho_4)^2 (1-\rho_3)(\rho_1+\rho_4)}
+ \frac{\rho_2\, \Big[(\tilde{t}^\gga+y^\gga)(y_\gga+\Pz_\gga) (z^\ga t_\ga)+i\delta(\T)
t_\gga(\tilde{t}^\gga-\Pz^\gga)\Big]}{(1-\rho_1-\rho_4)^2 (1-\rho_3)^2 (\rho_1+\rho_4)} \\
+\frac{\rho_3\, \big[i\delta(\T)-z_\gga (y^\gga+\tilde{t}^\gga)\big]
(t^\ga y_\ga)}{(1-\rho_1-\rho_4)^2 (1-\rho_3)^2}
+\frac{\big[-i\delta(\T)+z_\gga (y^\gga+\tilde{t}^\gga)\big]
(p_1 {}^\ga+p_2 {}^\ga)t_\ga}{(1-\rho_1-\rho_4)(1-\rho_3)^2} \Bigg\}\Ee\go CCC\,.
\end{multline}
Since by the partial integration procedure $
    \gx_1\gd(\gx_2)\equiv   {1}+  \gx_1(\p_{\gx_1}-\p_{\gx_2})$,
 \eq{G3=} yields  $G_{ 3}$ \eq{FRest3}.
 %, where a sum of $\gd(\T)$-proportional terms equals   $J_5$ \eq{Result4}.

% \renewcommand{\thesection}{\Alph{appendix}.}

\addtocounter{appendix}{1}
\renewcommand{\theequation}{\Alph{appendix}.\arabic{equation}}
\addtocounter{section}{1}
\setcounter{equation}{0}
 \addcontentsline{toc}{section}{\,\,\,\,\,\,\,Appendix    D: Details of the final step of the calculation}
 \renewcommand{\thesubsection}{\Alph{appendix}.\arabic{subsection}}
\setcounter{subsection}{0}
 \renewcommand{\thesection}{\Alph{appendix}}

\section*{Appendix D: Details of the final step of the calculation}
\label{AppE}

By virtue of  Eqs.~\eq{EEgx14=}-\eq{Egx=21e},   Eq.~\eq{SUM3} yields \bee&&
 %11111111111111111111111111111111
\label{SUM=} S_6
 =    + i \ff{   \eta^2  }{4 }\int d\Gamma \, \delta(\xi_3)       \\  \nn&&
 \Big\{
       %232323232323%232323232323%232323232323%232323232323
+(y+ \tilde{t}{} )^{\gga}z_{\gga}\ff{\gr_2}{(1- \gr_1 -\gr_4)(1-\gr_3) }{t}{}^\ga  (p_1+p_2 ){} _\ga
\gx_1  \ff{1-\gr_3-\gr_2}{(1-\gr_1-\gr_4 )(1-\gr_3 )^2}\,\, \Pz{}^\ga y_{\ga}
\\ \nn&&
+(y+ \tilde{t}{} )^{\gga}z_{\gga}\ff{\gr_2}{(1- \gr_1 -\gr_4)(1-\gr_3) }{t}{}^\ga  (p_1+p_2 ){} _\ga
 \gx_1 \ff{1-\gr_3-\gr_2}{(1-\gr_1-\gr_4 )(1-\gr_3 )^2}\big(y + \Pz{}  \big)^\ga\tilde{t}{}_{\ga}
\\ \nn&&
+(y+ \tilde{t}{} )^{\gga}z_{\gga}\ff{\gr_2}{(1- \gr_1 -\gr_4)(1-\gr_3) }{t}{}^\ga  (p_1+p_2 ){} _\ga
 (-)  \gx_1  \ff{\gr_2}{(1-\gr_1-\gr_4 )(1-\gr_3 )}\,(p_3{}^{\ga}+p_2{}^{\ga}) y_{\ga}
\\ \nn&&
-(y+ \tilde{t}{} )^{\gga}z_{\gga}\ff{\gr_2}{(1- \gr_1 -\gr_4)(1-\gr_3) }{t}{}^\ga  (p_1+p_2 ){} _\ga
%\\ \nn&&\times(-)
 \gx_1   \ff{ \gr_2}{(1-\gr_1-\gr_4 )(1-\gr_3)}(p_3{} +p_2{})^{\gb}\tilde{t}{}_{\gb}
+    %%%%%%%%%
\eee
\bee \nn&&
+(y+ \tilde{t}{} )^{\gga}z_{\gga}\ff{\gr_2}{(1- \gr_1 -\gr_4)(1-\gr_3) }{t}{}^\ga  (p_1+p_2 ){} _\ga
(-)      \ff{ \gr_1+\gr_4}{ (1-\gr_3 )^2}\,\,( (p{}_1 +p_2)  )^\ga y_{\ga}
 \\ \nn&&
+(y+ \tilde{t}{} )^{\gga}z_{\gga}\ff{\gr_2}{(1- \gr_1 -\gr_4)(1-\gr_3) }{t}{}^\ga  (p_1+p_2 ){} _\ga
 (-)      \ff{  \gr_4 }{ (1-\gr_3 )^2}\,\,{t}{} ^\ga y_{\ga}
\\ \nn&&
+(y+ \tilde{t}{} )^{\gga}z_{\gga}\ff{\gr_2}{(1- \gr_1 -\gr_4)(1-\gr_3) }{t}{}^\ga  (p_1+p_2 ){} _\ga
 (-)\ff{  \gr_1 }{(1-\gr_3)^2}      (p_1 +p_2 )^{\ga}{t}{}_{\ga}\\ \nn&&
%%%%%%%%%%%%%%%%%%%%%%%      \Big[  + \Ez(\overrightarrow{\p}_{\gr_2}-\overrightarrow{\p}_{\gr_3})E  \Big]+
  \\ \nn&&
  %%%%%%%%14141414141414141414141
   +             \ff{ \gr_2}{(1-\gr_1-\gr_4 )(1-\gr_3)^2}
     ( p_1{}+ p_2)^{\gga} \big(y + (1 -\gr_4 ){t}{}\big)_{\gga}  z_\ga(y+\tilde{t}{})^{\ga}
  (-)  \gx_1  \ff{\gr_2}{(1-\gr_1-\gr_4 )(1-\gr_3 )}\,\, {t}{}^\ga y_{\ga}
  \\ \nn&&
   +             \ff{ \gr_2}{(1-\gr_1-\gr_4 )(1-\gr_3)^2}
     ( p_1{}+ p_2)^{\gga} \big(y + (1 -\gr_4 ){t}{}\big)_{\gga}  z_\ga(y+\tilde{t}{})^{\ga}
\eee\bee \nn&&\times (-)  \gx_1   \ff{ \gr_2 }{(1-\gr_1-\gr_4 )(1-\gr_3)( \gr_1+\gr_4 ) }\big(
                           y^\ga+ \Pz{}^\ga \big) {t}{}_{\ga}
  \\ \nn&&
   +             \ff{ \gr_2}{(1-\gr_1-\gr_4 )(1-\gr_3)^2}
     ( p_1{}+ p_2)^{\gga} \big(y + (1 -\gr_4 ){t}{}\big)_{\gga}  z_\ga(y+\tilde{t}{})^{\ga}
  \ff{1}{ (1-\gr_3 )}\,\, {t}{}^\ga y_{\ga}\\ \nn&&
   +              \ff{ \gr_2}{(1-\gr_1-\gr_4 )(1-\gr_3)^2}
     ( p_1{}+ p_2)^{\gga} \big(y + (1 -\gr_4 ){t}{}\big)_{\gga}  z_\ga(y+\tilde{t}{})^{\ga}
    (-) \ff{  1 }{(1-\gr_3)}      (p_1 +p_2 )^{\ga}{t}{}_{\ga}
     %%%%%%%%%%%%%%%%%%%%%%%%%%%%%%%%%%%%%%     [ \overrightarrow{\p}_{\gr_4}-\overrightarrow{\p}_{\gr_1}] E\Big)
+  \eee\bee \nn&&
  %212121212121212121212121
  +       \ff{ \gr_2}{(1-\gr_1-\gr_4 )(1-\gr_3)^2}
       {t}{}^\gga \big(y +  (1-\gr_1-\gr_4 )( p_1{}  +p_2{} ){}\big)_\gga z_\ga(y+\tilde{t}{})^{\ga}
        \gx_1  \ff{\gr_3}{(1-\gr_1-\gr_4 )^2 }\,\,(  -   ( p_3+p_2)    )^\ga y_{\ga}
        \\\nn  &&
  +       \ff{ \gr_2}{(1-\gr_1-\gr_4 )(1-\gr_3)^2}
       {t}{}^\gga \big(y +  (1-\gr_1-\gr_4 )( p_1{}  +p_2{} ){}\big)_\gga z_\ga(y+\tilde{t}{})^{\ga}
           \gx_1   \ff{ \gr_3}{(1-\gr_1-\gr_4 )^2 }
\big(  (  -  ( p_3+p_2)    )^\ga \big)\tilde{t}{}_{\ga}
\\\nn  &&
  +       \ff{ \gr_2}{(1-\gr_1-\gr_4 )(1-\gr_3)^2}
       {t}{}^\gga \big(y +  (1-\gr_1-\gr_4 )( p_1{}  +p_2{} ){}\big)_\gga z_\ga(y+\tilde{t}{})^{\ga}
\\ \nn&&\times \gx_1  \ff{\gr_3\gr_4}{(1-\gr_1-\gr_4 ) (1-\gr_3 )(\gr_1+\gr_4)}\,\,   {t}{} ^\ga y_{\ga}
  \\\nn  &&
  +       \ff{ \gr_2}{(1-\gr_1-\gr_4 )(1-\gr_3)^2}
       {t}{}^\gga \big(y +  (1-\gr_1-\gr_4 )( p_1{}  +p_2{} ){}\big)_\gga z_\ga(y+\tilde{t}{})^{\ga}
  \gx_1  \ff{1}{ (1-\gr_3 )}\,(p_1 +p_2 )^{\ga} y_{\ga}
         \\\nn  &&
  +       \ff{ \gr_2}{(1-\gr_1-\gr_4 )(1-\gr_3)^2}
       {t}{}^\gga \big(y +  (1-\gr_1-\gr_4 )( p_1{}  +p_2{} ){}\big)_\gga z_\ga(y+\tilde{t}{})^{\ga}
    \gx_1   \ff{ 1}{ (1-\gr_3)}(p_1 +p_2 )^{\ga}\tilde{t}{}_{\ga}
        \eee\bee\nn  &&
  +       \ff{ \gr_2}{(1-\gr_1-\gr_4 )(1-\gr_3)^2}
       {t}{}^\gga \big(y +  (1-\gr_1-\gr_4 )( p_1{}  +p_2{} ){}\big)_\gga z_\ga(y+\tilde{t}{})^{\ga}
\\ \nn&&\times (-)  \gx_1   \ff{ \gr_2}{(1-\gr_1-\gr_4 )(1-\gr_3)} \ff{\gr_4}{(\gr_1+\gr_4)^2}
\big(y^\ga+ \Pz{}^\ga \big){t}{}_{\ga}
        \\\nn  &&
  +       \ff{ \gr_2}{(1-\gr_1-\gr_4 )(1-\gr_3)^2}
       {t}{}^\gga \big(y +  (1-\gr_1-\gr_4 )( p_1{}  +p_2{} ){}\big)_\gga z_\ga(y+\tilde{t}{})^{\ga}
(-)      \ff{ 1}{ (1-\gr_3 )}\,(p_1 +p_2 )^{\ga} y_{\ga}
        \\\nn  &&
  +       \ff{ \gr_2}{(1-\gr_1-\gr_4 )(1-\gr_3)^2}
       {t}{}^\gga \big(y +  (1-\gr_1-\gr_4 )( p_1{}  +p_2{} ){}\big)_\gga z_\ga(y+\tilde{t}{})^{\ga}
    (-)     \ff{  1 }{(1-\gr_3)}      (p_1 +p_2 )^{\ga}{t}{}_{\ga}
       + \eee \bee\nn  &&
         %33333333333333333333333333333
+  \gx_1\Big[
    \ff{ \gr_2\gr_2}{(1- \gr_1 -\gr_4)^3(1-\gr_3)^3( \gr_1+\gr_4 ) }
\big( y+  (1-\gr_1-\gr_4 )( p_1{}  +p_2{} )+  (1 -\gr_4 ){t}{}  \big)^\gga
\big( y +   \tilde{t}{}  \big)_{\gga} z_\ga {t}{}^{\ga}
  \\ \nn &&
-\ff{ \gr_3\gr_2}{(1- \gr_1 -\gr_4)^3(1-\gr_3)^3 }
 {\big( y   +    \tilde{t}{} \big)^\gga z_{\gga}   {t}{}^{\ga} y_{\ga}}
  \\ \nn &&
  +    \ff{ \gr_2}{(1- \gr_1 -\gr_4)^2(1-\gr_3)^3 } \big( y
   +    \tilde{t}{} \big)^\gga z_{\gga}       { (p_1 +p_2 )^{\ga}{t}{}_{\ga}}
  \Big]
   \big(y  + \Pz{}   \big)^\gb
 (y+\tilde{t}{})_{\gb}\Big\} \Ee\go CCC\,.
 \eee
   Terms from the \rhs of \eq{SUM=} with $\gx$-independent  pre-exponentials are considered  in Section \ref{NEgx},
   while those with $\gx_1$-proportional pre-exponentials are considered  in Section \ref{gx}.

\subsection{ $\gx_1$-independent  pre-exponentials }
\label{NEgx}
  Here we consider only pre-exponentials, omitting     for brevity integrals, integral measures  {\it etc} of \eq{SUM=}.
By virtue of the Schouten  identity taking into account that $\sum \gr_i=1$
Eq.~\eq{SUM=} yields
\bee\nn   && Integrand(S_6)\Big|_{\mod \gx}=(y+ \tilde{t}{} )^{\gn}z_{\gn}\Big\{%\\ \nn&&
-
\ff{\gr_2(\gr_1+\gr_4)}{(1- \gr_1 -\gr_4)(1-\gr_3)^3 }{t}{}^\ga  (p_1+p_2 ){} _\ga
     \,\,( (p{}_1 +p_2)  )^\ga y_{\ga}
\qquad \\ \nn&&
- \ff{\gr_2 \gr_4 }{(1- \gr_1 -\gr_4)(1-\gr_3)^3 }{t}{}^\ga  (p_1+p_2 ){} _\ga
        \,{t}{} ^\ga y_{\ga}
\\ \nn&&
- \ff{\gr_2\gr_1 }{(1- \gr_1 -\gr_4)(1-\gr_3)^3 }{t}{}^\ga  (p_1+p_2 ){} _\ga
       (p_1 +p_2 )^{\ga}{t}{}_{\ga}
 \\ \nn&&%%1414
    +             \ff{ \gr_2}{(1-\gr_1-\gr_4 )(1-\gr_3)^3}
     ( p_1{}+ p_2)^{\gga} \big(y + (1 -\gr_4 ){t}{}\big)_{\gga}
   \,\, {t}{}^\ga y_{\ga}\eee\bee \nn&&
   -           \ff{ \gr_2}{(1-\gr_1-\gr_4 )(1-\gr_3)^3}
     ( p_1{}+ p_2)^{\gga} \big(y + (1 -\gr_4 ){t}{}\big)_{\gga}
        (p_1 +p_2 )^{\ga}{t}{}_{\ga}
  \\ \nn&&
 -       \ff{ \gr_2}{(1-\gr_1-\gr_4 )(1-\gr_3)^3}
       {t}{}^\gga \big(y +  (1-\gr_1-\gr_4 )( p_1{}  +p_2{} ){}\big)_\gga     \,(p_1 +p_2 )^{\ga} y_{\ga}
        \\\nn  &&
  -      \ff{ \gr_2}{(1-\gr_1-\gr_4 )(1-\gr_3)^3}
       {t}{}^\gga \big(y +  (1-\gr_1-\gr_4 )( p_1{}  +p_2{} ){}\big)_\gga
               (p_1 +p_2 )^{\ga}{t}{}_{\ga}     \Big\}\Ee\go CCC =%\eee
%=========================================11111111111111
  %\bee
\nn  \\
\nn  &&=(y+ \tilde{t}{} )^{\gn}z_{\gn}\ff{\gr_2  }{(1- \gr_1 -\gr_4)(1-\gr_3)^3 }\Big\{%\\ \nn&&
   \gr_1{t}{}^\ga  (p_1+p_2 ){} _\ga
       (p_1 +p_2 )^{\ga}{t}{}_{\ga}
    +
     ( p_1{}+ p_2)^{\gga}  y _{\gga}
   \,\, {t}{}^\ga y_{\ga}
    \\ \nn&&
         -
     ( p_1{}+ p_2)^{\gga}  (1 -\gr_4 ){t}{}_{\gga}
        (p_1 +p_2 )^{\ga}{t}{}_{\ga}
 \\ \nn&&
  -
       {t}{}^\gga  y _\gga     \,(p_1 +p_2 )^{\ga} y_{\ga}
    -              {t}{}^\gga   (1-\gr_1-\gr_4 )( p_1{}  +p_2{} ) _\gga
               (p_1 +p_2 )^{\ga}{t}{}_{\ga}   \Big\}\Ee\go CCC
               \\ \nn %\eee\bee
     &&=(y+ \tilde{t}{} )^{\gn}z_{\gn}\ff{\gr_2  }{(1- \gr_1 -\gr_4)(1-\gr_3)^3 }\Big\{%\\ \nn&&
 - \gr_1{t}{}^\ga  (p_1+p_2 ){} _\ga
       (p_1{}^{\ga}+p_2{}^{\gb}){t}{}_{\gb}
 \\ \nn&&
           -     ( p_1{}+ p_2)^{\gga}  (1 -\gr_4 ){t}{}_{\gga}
        (p_1 +p_2 )^{\ga}{t}{}_{\ga}
    -
       {t}{}^\gga   (1-\gr_1-\gr_4 )( p_1{}  +p_2{} ) _\gga
               (p_1 +p_2 )^{\ga}{t}{}_{\ga}   \Big\}\Ee\go CCC\equiv 0 .\eee
\subsection{  $\gx_1$-proportional pre-exponentials}
\label{gx}
\bee\label{lostcohomo2}&&
%11111111111111111111111111111111
 S_6\, \Big|_{\gx_1  }
 =  J_7  +   i \ff{   \eta^2  }{4 }\int d\Gamma  \delta(\xi_3)     \\  \nn&&\Big\{
       %232323232323%232323232323%232323232323%232323232323
(y+ \tilde{t}{} )^{\gga}z_{\gga}\ff{\gr_2}{(1- \gr_1 -\gr_4)(1-\gr_3) }{t}{}^\ga  (p_1+p_2 ){} _\ga
\gx_1  \ff{1-\gr_3-\gr_2}{(1-\gr_1-\gr_4 )(1-\gr_3 )^2}\,\, \Pz{}^\ga y_{\ga}
\\ \nn&&
+(y+ \tilde{t}{} )^{\gga}z_{\gga}\ff{\gr_2}{(1- \gr_1 -\gr_4)(1-\gr_3) }{t}{}^\ga  (p_1+p_2 ){} _\ga
 \gx_1 \ff{1-\gr_3-\gr_2}{(1-\gr_1-\gr_4 )(1-\gr_3 )^2}\big(y + \Pz{}  \big)^\ga\tilde{t}{}_{\ga}
\\ \nn&&
-(y+ \tilde{t}{} )^{\gga}z_{\gga}\ff{\gr_2}{(1- \gr_1 -\gr_4)(1-\gr_3) }{t}{}^\ga  (p_1+p_2 ){} _\ga
   \gx_1  \ff{\gr_2}{(1-\gr_1-\gr_4 )(1-\gr_3 )}\,(p_3{}^{\ga}+p_2{}^{\ga}) y_{\ga}
\\ \nn&&
-(y+ \tilde{t}{} )^{\gga}z_{\gga}\ff{\gr_2}{(1- \gr_1 -\gr_4)(1-\gr_3) }{t}{}^\ga  (p_1+p_2 ){} _\ga
 \gx_1   \ff{ \gr_2}{(1-\gr_1-\gr_4 )(1-\gr_3)}(p_3{} +p_2{})^{\gb}\tilde{t}{}_{\gb}
%%%%%%%%%
   \\ \nn&&
 -            \ff{ \gr_2}{(1-\gr_1-\gr_4 )(1-\gr_3)^2}
     ( p_1{}+ p_2)^{\gga} \big(y + (1 -\gr_4 ){t}{}\big)_{\gga}  z_\ga(y+\tilde{t}{})^{\ga}
     \gx_1  \ff{\gr_2}{(1-\gr_1-\gr_4 )(1-\gr_3 )}\,\, {t}{}^\ga y_{\ga}
  \\ \nn&&
   -             \ff{ \gr_2}{(1-\gr_1-\gr_4 )(1-\gr_3)^2}
     ( p_1{}+ p_2)^{\gga} \big(y + (1 -\gr_4 ){t}{}\big)_{\gga}  z_\ga(y+\tilde{t}{})^{\ga}
    \gx_1   \ff{ \gr_2 }{(1-\gr_1-\gr_4 )(1-\gr_3)( \gr_1+\gr_4 ) }
  \\\nn  &&\times
  \big( y^\ga+ \Pz{}^\ga \big) {t}{}_{\ga}
   %\eee\bee
   \\ \ls\ls\ls\nn&&
  %212121212121212121212121
  +       \ff{ \gr_2}{(1-\gr_1-\gr_4 )(1-\gr_3)^2}
       {t}{}^\gga \big(y +  (1-\gr_1-\gr_4 )( p_1{}  +p_2{} ){}\big)_\gga z_\ga(y+\tilde{t}{})^{\ga}
        \gx_1  \ff{\gr_3}{(1-\gr_1-\gr_4 )^2 }\,\,(  -   ( p_3+p_2)    )^\ga y_{\ga}
        \\\nn  &&
  +       \ff{ \gr_2}{(1-\gr_1-\gr_4 )(1-\gr_3)^2}
       {t}{}^\gga \big(y +  (1-\gr_1-\gr_4 )( p_1{}  +p_2{} ){}\big)_\gga z_\ga(y+\tilde{t}{})^{\ga}
           \gx_1   \ff{ \gr_3}{(1-\gr_1-\gr_4 )^2 }
\big(  (  -  ( p_3+p_2)    )^\ga \big)\tilde{t}{}_{\ga}
\\\nn  &&
  +       \ff{ \gr_2}{(1-\gr_1-\gr_4 )(1-\gr_3)^2}
       {t}{}^\gga \big(y +  (1-\gr_1-\gr_4 )( p_1{}  +p_2{} ){}\big)_\gga z_\ga(y+\tilde{t}{})^{\ga}
 \\\nn  &&\times
  \gx_1  \ff{\gr_3\gr_4}{(1-\gr_1-\gr_4 ) (1-\gr_3 )(\gr_1+\gr_4)}\,\,   {t}{} ^\ga y_{\ga}
  \\\nn  &&
  +       \ff{ \gr_2}{(1-\gr_1-\gr_4 )(1-\gr_3)^2}
       {t}{}^\gga \big(y +  (1-\gr_1-\gr_4 )( p_1{}  +p_2{} ){}\big)_\gga z_\ga(y+\tilde{t}{})^{\ga}
  \gx_1  \ff{1}{ (1-\gr_3 )}\,(p_1 +p_2 )^{\ga} y_{\ga}
         \\\nn  &&
  +       \ff{ \gr_2}{(1-\gr_1-\gr_4 )(1-\gr_3)^2}
       {t}{}^\gga \big(y +  (1-\gr_1-\gr_4 )( p_1{}  +p_2{} ){}\big)_\gga z_\ga(y+\tilde{t}{})^{\ga}
    \gx_1   \ff{ 1}{ (1-\gr_3)}(p_1 +p_2 )^{\ga}\tilde{t}{}_{\ga}
        \\\nn  &&
  -       \ff{ \gr_2}{(1-\gr_1-\gr_4 )(1-\gr_3)^2}
       {t}{}^\gga \big(y +  (1-\gr_1-\gr_4 )( p_1{}  +p_2{} ){}\big)_\gga z_\ga(y+\tilde{t}{})^{\ga}
 \\\nn  &&\times
    \gx_1   \ff{ \gr_2}{(1-\gr_1-\gr_4 )(1-\gr_3)} \ff{\gr_4}{(\gr_1+\gr_4)^2}
\big(y^\ga+ \Pz{}^\ga \big){t}{}_{\ga}
         %\eee\bee
         \\ \nn  &&
         %33333333333333333333333333333
+  \gx_1\Big[  %%%%%%%%2222222222222
   \ff{ \gr_2\gr_2}{(1- \gr_1 -\gr_4)^3(1-\gr_3)^3( \gr_1+\gr_4 ) }
\big( y+  (1-\gr_1-\gr_4 )( p_1{}  +p_2{} )+  (1 -\gr_4 ){t}{}  \big)^\gga
\big( y +   \tilde{t}{}  \big)_{\gga}\\ \nn &&\times\Big\{
{t}{}^{\ga}\big(y  + \Pz{}   \big)_\ga     z^\gs (y+\tilde{t}{})_{\gs}
%+{t}{}^{\ga}(y+\tilde{t}{})_\ga \underline{ (i\gd(\T))} %
\Big\} %\\ \nn &&
-\ff{ \gr_3\gr_2}{(1- \gr_1 -\gr_4)^3(1-\gr_3)^3 }
 {\big( y   +    \tilde{t}{} \big)^\gga z_{\gga}   {t}{}^{\ga} y_{\ga}}
    \big(y  + \Pz{}   \big)^\gs
 (y+\tilde{t}{})_{\gs} \\ \nn &&
  +    \ff{ \gr_2}{(1- \gr_1 -\gr_4)^2(1-\gr_3)^3 } \big( y
   +      \tilde{t}{} \big)^\gga z_{\gga}     \big(y  + \Pz{}   \big)^\gs
 (y+\tilde{t}{})_{\gs}   { (p_1 +p_2 )^{\ga}{t}{}_{\ga}}
    \Big]
\Big\}\Ee\go CCC\q
 \eee
where   $J_7$ is the cohomology term    \eq{Result6}\,.
This yields %modulo  $J_7$
%%%%%%%%%%%%%%%%%%%%%%%%%%%%%%%%%%%%%%%%%%%%%%1111111111111111111111111111111111111
\bee\label{Endofgx}&&
 S_6\, \Big|_{\gx_1  }    \approx J_7 +   i \ff{   \eta^2  }{4 }\int d\Gamma \delta(\xi_3)
 \ff{\gr_2 }{(1- \gr_1 -\gr_4)^2(1-\gr_3)^2 }
      \\ \nn&&\gx_1(y+ \tilde{t}{} )^{\gga}z_{\gga}   \Big\{
 \ff{ (1-\gr_3-\gr_2)}{ (1-\gr_3)  }{t}{}^\ga  (p_1+p_2 ){} _\ga
    {\big(y + \Pz{}  \big)^\gb(y+ \tilde{t}{} )_{\gb}}
 \\ \nn&&
- \gr _2 {t}{}^\ga  (p_1+p_2 ){} _\ga
   (p_3{}^{\ga}+p_2{}^{\gb})(y+ \tilde{t}{} )_{\gb}
    -             \ff{\gr _2}{ (1-\gr_3) }
     ( p_1{}+ p_2)^{\gga} \big(y + (1 -\gr_4 ){t}{}\big)_{\gga}
    \,\, {t}{}^\ga y_{\ga}
  \\ \nn&&
  -        \ff{\gr _2}{ (1-\gr_3)  ( \gr_1+\gr_4 )}
     ( p_1{}+ p_2)^{\gga} \big(y + (1 -\gr_4 ){t}{}\big)_{\gga}
     \big(y+ \Pz{}  \big)^\ga {t}{}_{\ga}
 \\ \nn&&
    -     \ff{   \gr_3}{(1-\gr_1-\gr_4 )^2 }
       {t}{}^\gga \big(y +  (1-\gr_1-\gr_4 )( p_1{}  +p_2{} ){}\big)_\gga
  ( p_3+p_2)^\ga  (y+ \tilde{t}{} )_{\ga}
\\\nn  &&
  +       \ff{ \gr_3\gr_4}{ (1-\gr_3) (\gr_1+\gr_4)}
       {t}{}^\gga \big(y +  (1-\gr_1-\gr_4 )( p_1{}  +p_2{} ){}\big)_\gga
         \,   {t}{} ^\ga y_{\ga}
  \\\nn  &&
   +       \ff{  (1-\gr_1-\gr_4 )}{ (1-\gr_3) }
       {t}{}^\gga \big(y +  (1-\gr_1-\gr_4 )( p_1{}  +p_2{} ){}\big)_\gga
       (p_1 +p_2 )^{\ga}(y+ \tilde{t}{} )_{\ga}
        \\\nn  &&
  -     \ff{ \gr _2\gr_4}{ (1-\gr_3) (\gr_1+\gr_4)^2}
       {t}{}^\gga \big(y +  (1-\gr_1-\gr_4 )( p_1{}  +p_2{} ){}\big)_\gga
  \big(y^\ga+ \Pz{}^\ga \big){t}{}_{\ga}
    \\\nn  &&
     +\ff{  \gr_2}{(1- \gr_1 -\gr_4) (1-\gr_3) ( \gr_1+\gr_4 ) }
\big( y+    (1 -\gr_4 ){t}{}  \big)^\gga
\big( y +   \tilde{t}{}  \big)_{\gga}
{t}{}_{\ga}\big(y  + \Pz{}   \big)^\ga
\\ \nn &&
    +\ff{ \gr_2 }{ (1-\gr_3) ( \gr_1+\gr_4 ) }
  ( p_1{}  +p_2{} )^\gga
\big( y +   \tilde{t}{}  \big)_{\gga}
  {t}{}_{\ga}\big(y  + \Pz{}   \big)^\ga
  \\ \nn &&
-\ff{ \gr_3 }{(1- \gr_1 -\gr_4) (1-\gr_3)  }
     {t}{}^{\ga} y_{\ga}
    \big(y  + \Pz{}   \big)^\gs
 (y+\tilde{t}{})_{\gs} \\ \nn &&
  +    \ff{ 1}{ (1-\gr_3) }  \big(y  + \Pz{}   \big)^\gs
 (y+\tilde{t}{})_{\gs}    { (p_1 +p_2 )^{\ga}{t}{}_{\ga}}
\Big\}\Ee\go CCC\equiv J_7
 \eee
since, using  the Schouten  identity, one can see that
the pre-exponential of the integrand on the \rhs of   \eq{Endofgx}  equals   zero.
\addtocounter{appendix}{1}
\renewcommand{\theequation}{\Alph{appendix}.\arabic{equation}}
\addtocounter{section}{1} \setcounter{equation}{0}
 \addcontentsline{toc}{section}{\,\,\,\,\,\,\,Appendix E:   Useful  formulas}
 \renewcommand{\thesubsection}{\Alph{appendix}.\arabic{subsection}}

\section*{Appendix E:   Useful  formulas}
\label{AppG}

From \eq{Egx=}
 one has
 \bee\label{EEgx14=}&&\left(\frac{\p}{\p \rho_1}-\frac{\p}{\p \rho_4}\right) E=   i\Big\{
    \gx_1  \ff{\gr_2}{(1-\gr_1-\gr_4 )(1-\gr_3 )}\,\, {t}{}^\ga y_{\ga}
\\ \nn &&+  \gx_1   \ff{ \gr_2 }{(1-\gr_1-\gr_4 )(1-\gr_3)( \gr_1+\gr_4 ) }\big( y + \Pz{}\big)^\ga {t}{}_{\ga}
    \\ \nn &&
  +      \ff{  1 }{(1-\gr_3)}      (y+p_1{}^{\ga}+p_2{}^{\ga}){t}{}_{\ga}
 \Big\}E  \eee
   \bee\label{EEgx23}&&\left(\frac{\p}{\p \rho_2}-\frac{\p}{\p \rho_3}\right) E=  i  \Big\{
   \gx_1 \ff{1-\gr_3-\gr_2}{(1-\gr_1-\gr_4 )(1-\gr_3 )^2}
   \big( y + \Pz{}\big)^\ga(y+\tilde{t}{})_{\ga}
\\ \nn&& -  \gx_1  \ff{\gr_2}{(1-\gr_1-\gr_4 )(1-\gr_3 )}\,(p_3{}^{\ga}+p_2{}^{\ga}) (y+\tilde{t}{})_{\ga}
%%%%%%%%%%%%%%%%%%%%%%%%%%%%%
 %\\ \nn&&
 -      \ff{ \gr_1+\gr_4}{ (1-\gr_3 )^2}\,\,( (p{}_1 +p_2)  )^\ga y_{\ga}
\\ \nn && -      \ff{  \gr_4 }{ (1-\gr_3 )^2}\,\,{t}{} ^\ga y_{\ga}
%
 %  \\ \nn && % !!!!
  -      \ff{  \gr_1 }{(1-\gr_3)^2}      (p_1 +p_2 )^{\ga}{t}{}_{\ga}
 \Big\}
 E \,, \eee
\bee\label{Egx=21e}&&\left(\frac{\p}{\p \rho_2}-\frac{\p}{\p \rho_1}\right)  E = i  \Big\{
  \gx_1   \ff{ -\gr_3}{(1-\gr_1-\gr_4 )^2 }
 ( p_3+p_2  )^\ga  (y+\tilde{t}{})_{\ga}
\\ \nn&&
%%%%%%%%%%%%%%%
 +   \gx_1  \ff{\gr_3\gr_4}{(1-\gr_1-\gr_4 ) (1-\gr_3 )(\gr_1+\gr_4)}\,\,   {t}{} ^\ga y_{\ga}
%\\ \nn &&
%%%%%%%%%%%%%%%%
 +  \gx_1   \ff{ 1}{ (1-\gr_3)}(p_1 +p_2 )^{\ga}(y+\tilde{t}{})_{\ga}
\\ \nn &&-  \gx_1   \ff{ \gr_2}{(1-\gr_1-\gr_4 )(1-\gr_3)} \ff{\gr_4}{(\gr_1+\gr_4)^2}
\big( y + \Pz{}\big)^\ga{t}{}_{\ga}
 \\ \nn &&
\\ \nn&&
-     \ff{ 1}{ (1-\gr_3 )}\,(p_1 +p_2 )^{\ga} y_{\ga}
%   \\ \nn &&
-      \ff{  1 }{(1-\gr_3)}      (p_1 +p_2 )^{\ga}{t}{}_{\ga}
 \Big\}
   E \, . \eee

 \addcontentsline{toc}{section}{\,\,\,\,\,\,\,References}

\section*{}

\end{document}